\documentclass[a4paper,11pt]{article}

\pdfoutput=1

\usepackage{jheppub}
\usepackage{color}
\usepackage{graphicx}                   
\usepackage{amsmath}
\usepackage{amssymb}
\usepackage{amsfonts}
\usepackage{xspace}
\usepackage{verbatim}

\usepackage[small]{subfigure}
\usepackage{slashed}
\usepackage{mdwlist}
\usepackage[normalem]{ulem}
\usepackage[usenames,dvipsnames,svgnames,table]{xcolor}

\newcommand{\powR}{2}

\newcommand{\LL}{\text{LL}}
\newcommand{\NLL}{\text{NLL}}
\newcommand{\NNLL}{\text{N}^2\text{LL}}
\newcommand{\NNNLL}{\text{N}^3\text{LL}}

\newcommand{\SignOnPosBFKL}{-}

\allowdisplaybreaks[2]

  \newcount\hour \newcount\minute
  \hour=\time \divide \hour by 60 \minute=\time
  \newcommand{\todaytime}{\today \ -- \number\hour :\ifnum \minute<10 0\fi\number\minute}

\title{Soft Fragmentation on the Celestial Sphere}

\author[a]{Duff Neill,}
\author[b,c]{Felix Ringer}

\affiliation[a]{Theoretical Division, MS B283, Los Alamos National Laboratory, Los Alamos, NM 87545, USA}
\affiliation[b]{Physics Department, University of California, Berkeley, CA 94720, USA}
\affiliation[c]{Nuclear Science Division, Lawrence Berkeley National Laboratory, Berkeley, CA 94720, USA}
                                         
\emailAdd{duff.neill@gmail.com}
\emailAdd{fmringer@berkeley.edu}

\abstract{ We develop two approaches to the problem of soft fragmentation of hadrons in a gauge theory for high energy processes. The first approach directly adapts the standard resummation of the parton distribution function's anomalous dimension (that of twist-two local operators) in the forward scattering regime, using $k_T$-factorization and BFKL theory, to the case of the fragmentation function by exploiting the mapping between the dynamics of eikonal lines on transverse-plane to the celestial-sphere. Critically, to correctly resum the anomalous dimension of the fragmentation function under this mapping, one must pay careful attention to the role of regularization, despite the manifest collinear or infra-red finiteness of the BFKL equation. The anomalous dependence on energy in the celestial case, arising due to the mismatch of dimensionality between positions and angles, drives the differences between the space-like and time-like anomalous dimension of parton densities, even in a conformal theory. The second approach adapts an angular-ordered evolution equation, but working in $4-2\epsilon$ dimensions at all angles. The two approaches are united by demanding that the anomalous dimension in $4-2\epsilon$ dimensions for the parton distribution function determines the kernel for the angular-ordered evolution to all orders.}

\begin{document}
\maketitle
\section{Introduction}

A basic problem in high-energy scattering is that of fragmentation: how is the total energy of the scattering process divided amongst the final-state remnants? Fragmentation is a critical concern since it probes the dynamical properties of the theory. Not only must one know the bound states or resonances that the theory can generate, but their evolution and production during a scattering process. Even in a conformal theory, that lacks any hadronization to cutoff the infra-red dynamics, the question remains, but phrased slightly differently: how is energy distributed into calorimetry cells decorating the celestial sphere about the hard scattering origin \cite{Tkachov:1995kk}?  The latter question can be defined in terms of correlation functions of operators integrated along null directions, implicitly used in refs. \cite{Basham:1978bw,Basham:1978zq}, and more formally introduced in ref. \cite{Sveshnikov:1995vi}. These energy-correlation functions (EEC) have seen extensive use of in jet physics, for example, refs. \cite{Korchemsky:1999kt,Belitsky:2001ij,Lee:2006fn,Bauer:2008dt}, and have collinear limits that are controlled by fragmentation processes, refs. \cite{Dixon:2019uzg,Korchemsky:2019nzm}. Indeed, such correlation functions have received renewed attention in the context of conformal theory, either aiming at investigating genuine observables (infra-red finite) in $N=4$ super Yang-Mills, or more ambitiously formulating an operator product expansion that is quasi-local on the surface of the celestial sphere \cite{Hofman:2008ar,Belitsky:2013xxa,Kravchuk:2018htv,Henn:2019gkr,Kologlu:2019mfz}.

In a gauge theory respecting factorization, fragmentation is described by a product of a hard kernel or coefficient function, and the fragmentation function (FF). The hard kernel describes how the experimenter's probes couples to the underlying fields of the gauge theory, and the fragmentation function gives the number of hadrons carrying a specific fraction of the total energy as a function of the high energy scale one uses to probe the theory. The behavior of the fragmentation function as a function of this hard energy scale, and thus ultimately the scaling behavior of the cross-section, is controlled by the time-like Dokshitzer-Gribov-Lipatov-Alterelli-Parisi (DGLAP) anomalous dimension. The paradigmatic example requiring the FF is the case of single inclusive annihilation (SIA), where an electro-weak current at a time-like momentum scale decays to hadrons. The study of the soft region of fragmentation has a particular importance, since this is where the bulk of particles are produced in a high-energy collision dominated by quantum chromo-dynamics (QCD). In this region, the perturbative expansion for the anomalous dimension breaks down, necessitating resummation. The leading logarithmic (LL) resummation of the fragmentation spectrum in the soft region was accomplished long ago in ref. \cite{Mueller:1981ex,Bassetto:1982ma}. While this was then extended to next-to-leading logarithmic (NLL) accuracy (ref. \cite{Mueller:1982cq}), it was in a scheme with no straightforward relation to dimensional regularization, and thus could not be matched to standard calculations of the DGLAP anomalous dimensions. Instead, a so-called ``mixed leading log'' approximation was developed, with unclear systematics, but with phenomenological success in describing the distribution of hadrons in jets (for review, ref. \cite{Khoze:1996dn}). Actually addressing the resummation of the anomalous dimension in dimensional regularization took some time, and was initiated in ref. \cite{Albino:2011si}. However, a true break-through to higher logarithmic orders did not occur until refs. \cite{Vogt:2011jv,Kom:2012hd}, which introduced a novel scheme of recursion relations based on the cancellation of soft divergences, finally extending the resummation of the time-like DGLAP anomalous dimension \emph{and} the coefficient function for SIA to N$^2$LL accuracy (in double log counting, see Sec. \ref{sec:anom_dim_logs} below). This scheme has been implemented in fragmentation function fits in ref. \cite{Anderle:2016czy} and demonstrates an excellent description of the soft fragmentation data.

The counterpart to dynamical fragmentation is the static structure of the bound states of the theory, that is, how are the quantum numbers (spin, charge, invariant mass, etc.) divided amongst the constituents of the bound states, and whether this structure is independent of the particular scattering experiment used to probe it. Deep-inelastic scattering (DIS) provides the paradigmatic example of how to probe the structure of a bound state, where one investigates the electro-weak current inside the bound state at a specific space-like momentum scale. The cross-section here also factors into a product of functions, one describing the interactions of the hard space-like probe with the underlying gauge theory fields, and a function giving the distribution of the momentum of the bound state over its constituents, the parton distribution function (PDF). In contrast to the fragmentation function, the PDF can be defined entirely in terms of local-operators using the twist-expansion (ref. \cite{Christ:1972ms}), which ultimately allows one to give a complete non-perturbative way to calculate them using Monte Carlo Lattice Gauge theory \cite{Ji:2013dva,Izubuchi:2018srq}. This is in contrast to the FF, where seemingly one is forced to use non-local operators (like the EEC light-ray operators) to give a definition to the function.

The soft region of the DIS cross-section is of particular importance (like in SIA), since it probes the forward scattering region in the presence of strong interactions, and thus the mechanism by which QCD unitarizes the growth of the total cross-section with energy. The perturbative expansion for the anomalous dimension (the space-like Dokshitzer-Gribov-Lipatov-Alterelli-Parisi (DGLAP) anomalous dimension, or the anomalous dimension for twist-two operators of arbitrary spin) governing the scaling behavior of the DIS cross-section breaks down, necessitating a resummation. This resummation is now performed in the context of the Balitsky-Fadin-Kuraev-Lipatov (BFKL) equation (refs. \cite{Lipatov:1976zz,Kuraev:1976ge,Kuraev:1977fs, Balitsky:1978ic}). The BFKL approach represents a new factorization of the DIS cross-section, and consistency of this factorization with the factorization in terms of the PDF enables the resummation of the DIS cross-section in the soft region. 

Nonetheless, despite their physical differences, there has long been the desire to connect the PDF to the FF, since to relate DIS to SIA, one naively must simply interchange the observed hadronic state from the initial to the final state, and analytically continue the hard momentum initiating the process from a space-like to a time-like region. Pursuing this naive approach leads to the so-called Drell-Levy-Tan relations and the Gribov relations, refs. \cite{Drell:1969jm,Gribov:1972rt}, see also ref. \cite{Blumlein:2000wh}. Such straightforward relations are known to fail beyond leading order in gauge theories. Indeed, the established resummations of the time-like DGLAP anomalous dimension make no direct connection to the BFKL theory used to resum the space-like anomalous dimension. However, a more sophisticated relationship between the scaling properties of DIS and SIA appears to hold to all orders in perturbation theory, the so-called space-time reciprocity relations between the space-like and time-like DGLAP anomalous dimensions \cite{Moch:2004pa,Vogt:2004mw,Dokshitzer:2005bf,Basso:2006nk} of the PDF and the FF respectively.\footnote{To get a intuitive feeling about why it is perhaps bizarre to posit a relationship between DIS and SIA, one can simply compare their behavior as a function of the momentum fraction in the soft region. In the first, one has a power-law behavior of the cross-section with the small momentum fraction, and the other a skewed gaussian in log of the momentum fraction, see Figs. 18.2 and 19.4 of ref. \cite{Tanabashi:2018oca}. Both shapes are a result of the structure of the resummed space-like and time-like DGLAP anomalous dimensions, respectively.}

That such a reciprocity relation might exist is perhaps not surprising given the relatively recent discovery of a duality that exists between the soft dynamics found in forward scattering physics and the soft dynamics probed in jet physics. More precisely, the equations governing the scaling behavior of the effective theory (refs. \cite{McLerran:1993ka,McLerran:1993ni,McLerran:1994vd}) for forward scattering cross-section given by the BFKL equation and its extensions in the Balitsky-Kovchegov (BK) equation and the JIMWLK hierarchy (refs. \cite{Balitsky:1995ub,Kovchegov:1999yj,JalilianMarian:1996xn,JalilianMarian:1997gr,Iancu:2001ad}) are conformally equivalent to the Banfi-Marchesini-Smye (BMS) equation and its extensions (refs. \cite{Banfi:2002hw,Weigert:2003mm}) governing the energy distribution of entangled jet regions in exclusive jet cross-sections (ref. \cite{Dasgupta:2002bw}). The similarities between the sets of equations was noticed early on (ref. \cite{Weigert:2003mm}) and attempts to find the BFKL equation in the fragmentation processes of jets physics (ref. \cite{Marchesini:2003nh}) eventually led to the all-orders conformal map between the two regimes proposed in refs. \cite{Hatta:2008st,Avsar:2009yb}.\footnote{ For a more recent discussion see refs. \cite{Mueller:2018llt,Roy:2019hwr}.} This mapping has confirmation up to 2-loops in QCD (up to conformal anomalies) and 3-loops in $N=4$ super Yang-Mills (refs. \cite{Caron-Huot:2015bja,Caron-Huot:2016tzz}). Though the equations are equivalent, however, the initial-conditions and geometric restrictions are vastly different. Even so, one can still find analogs of saturation physics (for a short review, see ref. \cite{Kovchegov:2014kua}) which are responsible for unitarization governing the asymptotics of exclusive jet cross-sections (ref. \cite{Neill:2016stq,Larkoski:2016zzc}). However, no direct connection between the theory of soft fragmentation and the soft region of the PDF has been established.

The goal of this paper is to give such a direct connection. In particular, we wish to show the form of the BFKL equation that resums the \emph{time-like} DGLAP anomalous dimension, fully in dimensional regularization, establishing a direct connection between the theory which governs the soft momentum behaviour of the PDF and the FF. The key to understanding the BFKL equation in the time-like versus space-like case will be to always work in $4-2\epsilon$ dimensions until the very end of the calculation, since the only difference in the equations will be an anomalous dependence on the ordering variable (the logarithm of the momentum fraction). This is somewhat counter-intuitive, since the BFKL equation is manifestly finite as $\epsilon\rightarrow 0$, free of infra-red or collinear divergences. However, perturbation theory starts the initial condition for the BFKL equation with an ill-defined eigenfunction, so that regularization is actually imperative. Recognizing this need for such regularization was critical in the all orders resummation of the soft region of the DIS cross-section, first worked out in dimensional regularization in refs. \cite{Catani:1993ww,Catani:1994sq}. Here we argue the same is true for developing a BFKL theory for fragmentation.

Beyond developing a BFKL equation for time-like fragmentation, which we will call the ``celestial BFKL equation,'' we also give a version of the DGLAP equation using angular ordering, but working always in $4-2\epsilon$ dimensions. This allows us to recover results on the resummation of time-like DGLAP anomalous dimension (to N$^3$LL order in the minimal subtraction scheme) and the coefficient function (to N$^2$LL order) for SIA previously only derived using the recursion relation techniques of refs. \cite{Vogt:2011jv,Kom:2012hd}. This allows us to not only explicitly tie the resummation of the time-like anomalous dimension to the perturbative expansion of the space-like anomalous dimension, as expected from the reciprocity relation of refs. \cite{Moch:2004pa,Vogt:2004mw,Dokshitzer:2005bf,Basso:2006nk}, but we can further tie the resummation of the \emph{coefficient} function to the perturbative expansion of the space-like anomalous dimension. Finally, comparing the time-like BFKL equation to the angular-ordered DGLAP equations, we will point out some puzzles which will arise at three-loop order for the coefficient function.

The outline of the paper is as follows: first we review the factorization of the DIS and SIA cross-sections in terms of PDFs and FFs, giving the cross-section in terms of formal correlation functions of currents. We then review the all-orders definition of the FF and the PDF in dimensional regularization and minimal subtraction, where the functions appear as renormalization factors which absorb the infra-red divergences of the cross-section. These renormalization factors are wholly given by the anomalous dimensions. We then review the traditional BFKL equation and its application to resumming the PDF, and introduce the celestial BFKL equation. The celestial BFKL equation recovers leading log results (in the time-like counting), and partial results at all subleading logarithmic orders, but suffers from lack of manifest log counting. Inspired by the variables used in the celestial BFKL equation, and the structure of its iterative solution, we are then lead introduce the angular-ordered DGLAP evolution equation in dimensional regularization, and developing its consequences. It will enjoy manifest log counting, but points to a puzzle arising at three-loops/N$^3$LL logarithmic order in the coefficient function resummation. However, it will reproduce all known results for pure Yang-Mills theory found in the literature. We postulate a specific relationship between the $d$-dimensional space-like  DGLAP anomalous dimension, and the kernel of the angular-ordered DGLAP evolution, providing an extension of the reciprocity equation between space-like and time-like anomalous dimensions to include the resummation of the coefficient functions. We then conclude with proposals for future directions and unresolved questions.

\section{Review of Factorization with PDFs and FFs}

A standard observable for fragmentation is the inclusive cross-section for the process $a+b\rightarrow h + X$:\footnote{Since we do not care about the type of hadron that is fragmented, we will not include any subscript $h$ on our fragmentation form-factors or functions, to declutter notation.}
\begin{align}
  \frac{1}{\sigma_0}\frac{d\sigma}{dx}&=D\Big(x,\Lambda^2,Q^2\Big)\,,\\
    x&=\frac{2q\cdot k}{Q^2}\,.
\end{align}
where $h$ is a final state particle with momentum $k$ carrying energy fraction $x$, $a$ and $b$ are the intial scattering states with total momentum $q$ and $q^2=Q^2>0$, and $X$ is the rest of the final state, about which we are indifferent. For single inclusive annihilation (SIA) we take $a$ and $b$ to be an electron-positron pair, so that radiative corrections to the initial state can be ignored, and we can focus on the fragmentation process. $\Lambda$ can be considered the invariant mass-scale of final state particles (i.e., $h$), and $\Lambda\ll Q$. The function $D$ is determined as follows: one counts the number of particles in each event with energy fraction $x$ that fall into a momentum bin of size $dx$, then averages over events:
\begin{align}
  dN(x,Q)&=D(x,Q) dx\,.
\end{align}
In a gauge theory like QCD, the fragmentation cross-section enjoys a factorization theorem, as defined to all orders in perturbation theory in ref. \cite{Collins:1981uw}:
\begin{align}\label{eq:basic_factorization_FF}
  x D\Big(x,\Lambda^2,Q^2\Big)&=\sum_{a}\int_{x}^{1}\frac{dz}{z} \Big[\frac{x}{z}d_{a}\Big(\frac{x}{z},\Lambda^2,\mu^2\Big)\Big] zC^T_{a}\Big(z,Q^2,\mu^2\Big)+O\Big(\frac{\Lambda}{Q}\Big)\,.
\end{align}
All infra-red physics is found in $d$, the universal fragmentation function, containing $\Lambda$ the infra-red mass scale of the theory, $x$ is the energy fraction carried by the hadron of the event, while $C^T$ is the coefficient function describing the high energy process in the scattering cross-section. The convolution variable $z$ is the energy fraction of the parton exiting the high-energy scattering which will act as the parent of the observed particle. The label $a$ denotes the flavor of this intermediate state that exits the hard process. Further, the ratio $x/z$ is the energy fraction of the hadron with respect to the parton that fragments it. The exact decomposition between the coefficient function $C^T$ and the fragmentation function $d$ is regularization scheme dependent, and introduces an arbitrary factorization scale $\mu$ where the infra-red and ultra-violet modes are separated. 

Since the cross-section does not depend on the factorization of infra-red and ultra-violet processes, $D$ is independent of $\mu$, so that we can write a renormalization group equation (RGE) for $d$. This RGE is most easily expressed in moment space:
\begin{align}
\bar{d}_{a}(n,\Lambda^2,\mu^2)&=\int_{0}^{1}\frac{dz}{z}z^{n}\Big( z d_{a}(z,\Lambda^2,\mu^2)\Big)\,,\\
\mu^2\frac{d}{d\mu^2}\bar{d}_a(n,\Lambda^2,\mu^2)&=\sum_{b}\gamma^T_{ab}(n)\bar{d}_{b}(n,\Lambda^2,\mu^2)\,.
\end{align}  
This renormalization group equation will control the behavior of the cross-section as a function of $Q$, by integrating it from a boundary condition at $\Lambda$ up to the scale $Q$.

While fragmentation probes the dynamics of the theory, processes like deeply-inelastic scattering (DIS) probe how momentum, charge, spin, etc., of a bound state of the theory are distributed over the underlying constituents of the theory. Here one sends in a probe to scatter off of the bound state, and measures the cross-section for how it scatters. A classic example is electron-proton scattering, where one wants the cross-section for $e+p\rightarrow e+X$, $X$ being an otherwise arbitrary final state we are indifferent to:
\begin{align}
\frac{1}{\sigma}\frac{d\sigma}{ dx dQ^2}&=F(x,\Lambda^2,Q^2)\,.
\end{align}
We measure the momentum difference of the electron $q=P_{e_f}-P_{e_i}$, which is a space-like momentum transfer: $q^2=-Q^2$, $Q^2>0$. $\Lambda$ is again the infra-red scale of the theory. We let $P$ be the momentum of the proton, and then:
\begin{align}
x=\frac{Q^2}{2P\cdot q}\,.
\end{align}
Again, we have a factorization theorem within a gauge theory for the DIS process:
\begin{align}\label{eq:basic_factorization_PDF}
xF(x,\Lambda^2,Q^2)&=\sum_{a}\int_{x}^{1}\frac{dz}{z}\Big[\frac{x}{z}f_{a}\Big(\frac{x}{z},\Lambda^2;\mu^2\Big)\Big]z C^{S}_{a}\Big(z,Q^2;\mu^2\Big)+O\Big(\frac{\Lambda}{Q}\Big)
\end{align}  
$C^S$ encodes the short-distance interactions of the high-energy process, while $f$, the parton distribution function (PDF) (given a field theory definition in ref. \cite{Collins:1981uw}), gives the distribution to find a parton carrying momentum fraction $x$ of the proton's momentum $P$ that feeds into the high energy interaction. Again the exact decomposition between the coefficient function $C^S$ and the PDF $f$ is regularization scheme dependent, and introduces an arbitrary factorization scale $\mu$ where the infra-red and ultra-violet modes are separated.

As a scattering process, DIS may not seem of much use to a conformal theory, since we would have no bound states to probe. However, since the scattering process involves a space-like momentum transfer, and we are agnostic about the structure of the final state apart from the probe, we can use a dispersion relation to connect the DIS process to expectation values of local operators, the so-called twist expansion. The scaling behavior of such operators are the well-defined ``observables'' of a conformal theory. In particular, if we take the moment transform of $f$:
\begin{align}
  \bar{f}_{a}(n,\Lambda^2,\mu^2)&=\int_{0}^{1}\frac{dx}{x}x^{n}\Big( xf_{a}(x,\Lambda^2,\mu^2)\Big)\,,\\
  \mu^2\frac{d}{d\mu^2}\bar{f}_{a}(n,\Lambda^2,\mu^2)&=\sum_{b}\gamma^{S}_{ba}(n)\,\bar{f}_{b}(n,\Lambda^2,\mu^2)\,.
\end{align}  
The anomalous dimension $\gamma^{S}(n)$ is also the anomalous dimension of local operators with scaling dimension $n+2$ and spin $n$.

In what follows, we will call the observable functions $D$ and $F$ the SIA and DIS form factors. In general, the form factors can be expressed in terms of matrix elements of current operators $\hat{J}$:
\begin{align}
D\Big(x,\Lambda^2,Q^2\Big)&=\sum_{X,a,\sigma}\int d^dre^{iq\cdot r}\langle 0|\hat{J}(r)|X,k^{a\sigma}\rangle\langle X,k^{a\sigma}|\hat{J}(0)|0\rangle\,,\\
F\Big(x,\Lambda^2,Q^2\Big)&=\sum_{X}\int d^dre^{iq\cdot r}\langle P|\hat{J}(r)|X\rangle\langle X|\hat{J}(0)|P\rangle\,,
\end{align}
Note that the sum over states in both theories is at asymptotically late times, though the current insertions are at finite times. In the DIS form factor, we may remove the sum over states, and use a dispersion relation to get to a time-ordered product whose expression in terms of local operators is well-defined. In SIA, this is not possible.

To simplify the discussion, we will focus on pure Yang-Mills, with no quarks or colored scalars. This will allow us to drop flavor indicies in the following. We can probe the theory through a higher dimensional operator $\phi F^{A}_{\mu\nu}F^{A \mu\nu}$, where $\phi$ is a colorless scalar field and $F^{A\mu\nu}$ is the field strength tensor. Then in the above form factors, we take the current $J$ to be:
\begin{align}
J&= F^{A}_{\mu\nu}F^{A \mu\nu}\,.
\end{align}  
Further, we can appropriately normalize the form factors such that at tree-level:
\begin{align}
D\Big(x,\Lambda^2,Q^2\Big)=F\Big(x,\Lambda^2,Q^2\Big)=\delta(1-x)\,.
\end{align}  

\subsection{Factorization in Dimensional Regularization}
We wish to examine eqs. \eqref{eq:basic_factorization_FF} and \eqref{eq:basic_factorization_PDF} when we use dimensional regularization. In perturbation theory, both the SIA and DIS form factors are ill-defined, being infra-red divergent. The most convenient regularization to fix this problem is dimensional regularization, where we continue to $d=4-2\epsilon$ space-time dimensions. The form factor can then be calculated in perturbation theory, and we can give all-orders definitions to the FF and the PDF in terms of an anomalous dimension and the $\epsilon$ parameter used to continue the dimension of space-time. That is, the renormalized functions are the infra-red renormalization factors for the hard matching coefficient. In what follows, to avoid convolutions, we define the moment space transform:
\begin{align}
  \bar{g}(n)&=\int_0^{1}\frac{dx}{x}x^n(xg(x))\,,\\
  xg(x)&=\int\displaylimits_{c-i\infty}^{c+i\infty}\frac{dn}{2\pi i} x^{-n}\bar{g}(n)
\end{align}
So the factorization becomes:
\begin{align}
  \bar D\Big(n,\Lambda^2,Q^2\Big)&=\bar{d}\Big(n,\Lambda^2,\mu^2\Big) \bar{C}^T\Big(n,Q^2,\mu^2\Big)+O\Big(\frac{\Lambda}{Q}\Big)\,.
\end{align}  
In dimensional regularization, the higher order corrections to the \emph{bare} fragmentation function are given by scaleless integrals, which are set to zero, and we have:
\begin{align}
\bar{d}\Big(n,\Lambda^2,\mu^2\Big)&\rightarrow 1\,.
\end{align}  
This is true for the \emph{bare} function only. In truth, once we renormalize the ultra-violet divergences, the fragmentation function is completely determined by the infra-red divergences of the form factor:
\begin{align}\label{eq:Factor_IR_Divergences}
  \bar D\Big(n,Q^2\Big)&=\text{exp}\Big(\int\displaylimits_{0}^{\alpha_s(\mu^2)}\frac{d\alpha}{\beta(\alpha,\epsilon)}\gamma^T(\alpha,n)\Big)\,C^T\Big(n,Q^2,\mu^2\Big).
\end{align}
The function $\beta(\alpha_s,\epsilon)$ is the beta function of the theory in $4-2\epsilon$ dimensions and the function $\gamma^T(\alpha,n)$ is the time-like DGLAP anomalous dimension, acting as the kernel of the renormalization-group equation:
\begin{align}\label{eq:RG_gamma_T}
  \mu^2\frac{d}{d\mu^2}Z^{-1}_T&=\gamma^T Z^{-1}_T\,,\\
  Z^{-1}_T(n,\mu^2,\epsilon)&=\text{exp}\Big(\int\displaylimits_{0}^{\alpha_s(\mu^2)}\frac{d\alpha}{\beta(\alpha,\epsilon)}\gamma^T(\alpha,n)\Big),\\
  \mu^2\frac{d\alpha_s}{d\mu^2}&=\beta(\alpha_s,\epsilon),\label{eq:beta_function_def}\\
  \beta(\alpha_s,\epsilon)&=-\alpha_s\Big(\epsilon+\beta_0\frac{\alpha_sC_A}{\pi}+\beta_1\Big(\frac{\alpha_sC_A}{\pi}\Big)^2+...\Big),\\
  \beta_0&=\frac{11}{12},\,\,\,  \beta_1=\frac{17}{24}.
\end{align}  
The \emph{renormalized} fragmentation function within dimensional regularization (in the $\overline{MS}$-scheme) is simply the inverse renormalization factor $Z^{-1}$:
\begin{align}
  \bar{d}\Big(n,\Lambda^2,\mu^2\Big)\Big|_{\text{bare}}= 1 = Z_T(n,\mu^2,\epsilon)Z^{-1}_T(n,\mu^2,\epsilon)\,,\\
  \bar{d}\Big(n,\Lambda^2,\mu^2\Big)\Big|_{\text{renorm.}}= Z^{-1}_T(n,\mu^2,\epsilon)\,.
\end{align}  
Instead of a genuine infra-red scale $\Lambda$, we have the mass-scale $\mu$ of dimensional regularization which is now tied to the implicit renormalization scale of the coupling constant $\alpha_s$. Similarly for the renormalized PDF, we have the result:
\begin{align}\label{eq:pdf_in_dim_reg}
  \bar{f}\Big(n,\Lambda^2,\mu^2\Big)=Z^{-1}_S(n,\mu^2,\epsilon)=\text{exp}\Big(\int\displaylimits_{0}^{\alpha_s(\mu^2)}\frac{d\alpha}{\beta(\alpha,\epsilon)}\gamma^S(\alpha,n)\Big)\,,
\end{align}
where $\gamma^S$ is the space-like DGLAP anomalous dimension.

Working in dimensional regularization, we never see the scale $\Lambda$ directly. Indirectly, the location of the landau pole for the beta-function acts as a surrogate for the infra-red scale in dimensional regularization, which through RG running can be traded for the value of the renormalized coupling at some fixed UV scale. Henceforth, we shall drop explicit dependence on $\Lambda^2$, since we are interested in the dependence on the scale $\mu$ in what follows.

\subsection{Structure of Anomalous Dimensions and Resummed Perturbation Theory}\label{sec:anom_dim_logs}
Since we are interested in the resummation of the cross-section as the momentum fraction becomes small, we review the logarithmic structure for the anomalous dimensions in this region. In both the space-like and time-like DGLAP anomalous dimensions,\footnote{The space-like anomalous dimension has been calculated to three-loop order in ref. \cite{Moch:2004pa,Vogt:2004mw}, and while there are results for the time-like at the same order using reciprocity relations and sum rules, see refs. \cite{Moch:2007tx,Gituliar:2015pra}, no full direct calculation has been performed.} at fixed order in perturbation theory, the anomalous dimensions become dominated by the logarithm of the momentum fraction:
\begin{align}
  \lim_{x\rightarrow 0}\gamma^S(x)&=\sum_{\ell=0}^{\infty}\sum_{j=0}^{\ell}\Big(\frac{\alpha_sC_A}{\pi}\Big)^{1+\ell}\tilde \gamma^{S}_{\ell,j}\frac{\text{ln}^{j}\frac{1}{x}}{x}\,,\\
  \lim_{x\rightarrow 0}\gamma^T(x)&=\sum_{\ell=0}^{\infty}\sum_{j=0}^{2\ell}\Big(\frac{\alpha_sC_A}{\pi}\Big)^{1+\ell}\tilde \gamma^{T}_{\ell,j}\frac{\text{ln}^{j}\frac{1}{x}}{x}\,.
\end{align}  
In Mellin space, the logarithms of $x$ are translated into poles via the mapping:
\begin{align}
\int_{0}^{1}\frac{dx}{x}x^{n+1}\Big(\frac{1}{x}\text{ln}^{k}\frac{1}{x}\Big)&=\frac{\Gamma(1+k)}{n^{1+k}}\,.
\end{align}
Thus the soft region in $x$ maps to the moment $n$ going to zero, so that the soft region of the anomalous dimension is given by the laurent expansion of the anomalous dimensions at $n=0$:
\begin{align}\label{eq:laurent_space}
\gamma^S(n)&=\int_{0}^{1}\frac{dx}{x}x^{n}\Big(x\gamma^S(x)\Big)=\sum_{\ell=0}^{\infty}\sum_{j=-\ell-1}^{\infty}\Big(\frac{\alpha_sC_A}{\pi}\Big)^{1+\ell}n^{j}\gamma^{S}_{\ell,j}\,,\\
\label{eq:laurent_time}\gamma^T(n)&=\int_{0}^{1}\frac{dx}{x}x^{n}\Big(x\gamma^T(x)\Big)=\sum_{\ell=0}^{\infty}\sum_{j=-2\ell-2}^{\infty}\Big(\frac{\alpha_sC_A}{\pi}\Big)^{1+\ell}n^{j}\gamma^{T}_{\ell,j}\,.
\end{align}  
In general, one must be careful in relating the momentum and moment space soft regions, since the anomalous dimensions have plus-distributions in $1-x$, so that the laurent series in $n$ receives nontrivial contributions from the whole integration range in the moment transform. $\gamma^S$ has a single logarithmic behavior, meaning that each additional power of $\alpha_s$ in the laurent expansion comes with at most one additional logarithm, or a pole in $n$ one order higher. In contrast $\gamma^T$ has a double logarithmic behavior, meaning that each additional power of $\alpha_s$ comes with two additional logarithms, that is, a pole two orders higher. Then the resummed perturbation theory for both in the limit has the power counting:
\begin{align}\label{eq:log_count_space}
\text{Space-Like: }&  \alpha_s\sim n\ll 1,\quad\gamma^S(n)=\gamma^S_{\LL}\Big(\frac{\alpha_s}{n}\Big)+n\gamma^S_{\NLL}\Big(\frac{\alpha_s}{n}\Big)+n^2\gamma^S_{\NNLL}\Big(\frac{\alpha_s}{n}\Big)+...\\
\label{eq:log_count_time}\text{Time-Like: }&   \alpha_s\sim n^2\ll 1,\quad\gamma^T(n)=n\Bigg(\gamma^T_{\LL}\Big(\frac{\alpha_s}{n^2}\Big)+n\gamma^T_{\NLL}\Big(\frac{\alpha_s}{n^2}\Big)+n^2\gamma^T_{\NNLL}\Big(\frac{\alpha_s}{n^2}\Big)\Bigg)+...
\end{align}
where $\gamma_{\text{N}^k\text{LL}}^{S/T}$ is the k-th correction to the leading logarithmic resummation of the anomalous dimension. Note that the time-like anomalous dimension has an overall factor of $n$, this is due to the leading-order anomalous dimension having no logarithms in momentum-space, only a simple pole at $x=0$.

\section{BFKL in $4-2\epsilon$ Dimensions}\label{sec:space_like_BFKL}
We briefly recall the relationship between collinear factorization of the DIS form factor and the BFKL factorization (or ``$k_T$'' factorization) as derived in ref. \cite{Catani:1994sq}. Going to moment space, the form factor can be written as:
\begin{align}
  xF\Big(x,Q^2\Big)&=\int\displaylimits_{c-i\infty}^{c+i\infty}\frac{dn}{2\pi i} x^{-n}C^S\Big(n,Q^2;\mu^2\Big)f\Big(n;\mu^2\Big)\nonumber\\
  &=\int\displaylimits_{c-i\infty}^{c+i\infty}\frac{dn}{2\pi i} x^{-n}\int\frac{d^{2}\vec{k}_\perp}{2\pi} h^S\Big(n,Q^2;\vec{k}_\perp^{\,2}\Big)\bar{\mathcal{F}}\Big(n;\vec{k}_\perp^{\,2}\Big)\label{eq:BFKL_factorization}
\end{align}  
On the first line, we have recalled the factorization in terms of PDFs, dropping the sum over flavors. In the second line, we have written the factorization in terms of the BFKL impact factor (or Green's function). The intuitive interpretation of this impact factor is the parton density interacting with a pomeron carrying transverse momentum $\vec{k}_\perp$. Both factorizations have a process dependent part (the $C^S$ and the $h^S$ functions), and process independent functions ($f$ and $\mathcal{F}$) which are universal, but scheme dependent. Thus the same functions $\mathcal{F}$ will appear in high-energy resummations of the Drell-Yan process, Higgs production, or DIS, while $h^S$ changes for each of these (see for instance refs. \cite{Marzani:2008uh,Caola:2010kv}). Even within DIS, we will have distinct $h^S$ for the different polarization structures of the form factor. The function $\mathcal{F}$ will obey the BFKL equation, which will determine its $n$-dependence.

Consistency between the two factorizations means that the impact factor itself will factorize onto the PDF. At leading-logarithmic level in moment-space, this relationship is given as:
\begin{align}\label{eq:impact_factor_collinear_factorization}
\bar{\mathcal{F}}\Big(n,\vec{k}_\perp\Big)&=\gamma^S(\alpha_s,n) \frac{R^S(\alpha_s,n)}{\vec{k}^{\,2}_\perp}\Big(\frac{\vec{k}_{\perp}^{\,2}}{\mu^2}\Big)^{\gamma^S(\alpha_s,n)}\text{exp}\Big(\int\displaylimits_{0}^{\alpha_s(\mu^2)}\frac{d\alpha}{\beta(\alpha,\epsilon)}\gamma^S(\alpha,n)\Big)\,.
\end{align}  
The factor $R^{S}(n)$ is universal for any high-energy factorization using the impact factor, but is explicitly tied to the scheme used to define the anomalous dimension $\gamma^S$. The final exponential factor we can recognize as the infra-red divergences of the PDF in dimensional regularization, eq. \eqref{eq:pdf_in_dim_reg}. Ultimately the factor $R^{S}(n)$ arises from the fact that the correct initial condition for the BFKL equation is ill-defined in strictly 4 dimensions, necessitating a regularization of the BFKL equation. Dimensional regularization is thus natural, since this is the procedure by which we want to define the PDF anyways. One uses the BFKL equation in $4-2\epsilon$ dimensions to resum the $n$ dependence in both $R^S$ and $\gamma^S$.

The actual determination of the collinear factorization of the impact factor proceeds as follows. The momentum space BFKL equation in $4-2\epsilon$ dimensions is
{\small\begin{align}\label{eq:BFKL}
    x\frac{d}{dx}\mathcal{F}\Big(x,\vec{k}_\perp\Big)&=-\mathcal{F}\Big(x,\vec{k}_\perp\Big)\nonumber\\
    &\qquad-2\frac{\alpha_s C_A}{\pi}(4\pi e^{-\gamma_E})^{-\epsilon}\mu^{2\epsilon}\int\frac{d^{2-2\epsilon}\vec{q}_\perp}{(2\pi)^{1-2\epsilon}}\Big\{\frac{\mathcal{F}\Big(x,\vec{q}_\perp\Big)}{(\vec{q}_\perp-\vec{k}_\perp)^2}-\frac{\vec{k}_\perp^{\,2}}{2\vec{q}_\perp^{\,2}(\vec{q}_\perp-\vec{k}_\perp)^{\,2}}\mathcal{F}\Big(x,\vec{k}_\perp\Big)\Big\}\,,\\
  &=-\mathcal{F}\Big(x,\vec{k}_\perp\Big)-\frac{\alpha_s C_A}{\pi} K\otimes_{\perp}\mathcal{F}\Big(x,\vec{k}_\perp\Big)\,.
\end{align}}
We then take the moment transform to get:
\begin{align}\label{eq:BFKL_moment}
\bar{ \mathcal{F}}\Big(n,\vec{k}_\perp\Big)&=\frac{1}{n}c(\vec{k}_\perp)+\frac{\alpha_s C_A}{\pi\,n} K\otimes_{\perp}\bar{ \mathcal{F}}\Big(n,\vec{k}_\perp\Big)\,.
\end{align}
The function $c(\vec{k}_\perp)$ is the boundary condition to the BFKL equation now in moment space, taking for instance:\footnote{Other initial conditions, like $c(\vec{k}_\perp)=\delta^{(2-2\epsilon)}(\vec{k}_\perp)$, are possible, and useful for deriving the Green's function of the BFKL equation in $4-2\epsilon$ dimensions, but the conclusions are the same as the power-law ansatz used here.}
\begin{align}\label{eq:perturbation_theory}
c(\vec{k}_\perp)&=\frac{\alpha_sC_A}{\pi\,\vec{k}_\perp^{\,2}}\Big(\frac{\vec{k}_\perp^{\,2}}{\mu^2}\Big)^{\delta}\,.
\end{align}  
We use the fact that then the action of the BFKL kernel on such a power-law function has the form:
\begin{align}
K\otimes_{\perp}\Bigg(\frac{1}{\vec{k}_\perp^{\,2}}\Big(\frac{\vec{k}_\perp^{\,2}}{\mu^2}\Big)^{\delta}\Bigg)&=\frac{I(\delta,\epsilon)}{\vec{k}_\perp^{\,2}}\Big(\frac{\vec{k}_\perp^{\,2}}{\mu^2}\Big)^{\delta-\epsilon}
\end{align}  
This can be deduced from simple dimensional analysis, and we explicitly calculate the function $I(\delta,\epsilon)$ to be:
\begin{align}
I(\delta,\epsilon)=\frac{1}{\epsilon}e^{\epsilon \gamma_E}\Gamma(1-\epsilon)\Bigg(\frac{\Gamma(1-\epsilon)\Gamma(1+\epsilon)}{\Gamma(1-2\epsilon)}-\frac{\Gamma(\delta-\epsilon)\Gamma(1-\delta+\epsilon)}{\Gamma(1-\delta)\Gamma(\delta-2\epsilon)}\Bigg)\label{eq:I_def}
\end{align}  
We can expand iteratively eq. \eqref{eq:BFKL_moment}, writing:
\begin{align}
  \bar{\mathcal{F}}\Big(n,\vec{k}_\perp\Big)&=\frac{\alpha_sC_A}{\pi\,n\vec{k}_\perp^{\,2}}\Big(\frac{\vec{k}_\perp^{\,2}}{\mu^2}\Big)^{\delta}\sum_{\ell=0}^{\infty}c_{\ell}(\delta,\epsilon)\Big[\frac{\alpha_s C_A}{\pi n}\Big(\frac{\mu^2}{\vec{k}_\perp^{\,2}}\Big)^{\epsilon}\Big]^{\ell}\,,\\
  c_{0}&=1,\\
  c_{\ell+1}(\delta,\epsilon)&=I\Big(\delta-\ell\epsilon,\epsilon\Big)c_{\ell}(\delta,\epsilon)\,.\label{eq:c_def}
\end{align}
Setting $\delta=0$, then in the limit that $\epsilon\rightarrow 0$, we calculate:
{\small\begin{align}
   \bar{\mathcal{F}}&=\gamma^S(\alpha_s,n) R^S(\alpha_s,n)\Big(\frac{\vec{k}_{\perp}^{\,2}}{\mu^2}\Big)^{\gamma^S(\alpha_s,n)}\text{exp}\Big(\int_{0}^{\alpha_s}\frac{d\alpha}{\beta(\alpha,\epsilon)}\gamma^S(\alpha,n)\Big)\,,\\
  \gamma^S(\alpha_s,n)&=\frac{\alpha_s C_A}{\pi n}+2\zeta_3\Big(\frac{\alpha_s C_A}{\pi n}\Big)^4+2\zeta_5\Big(\frac{\alpha_s C_A}{\pi n}\Big)^6+...\,,\\
  R^S(\alpha_s,n)&=1+\frac{8}{3}\zeta_3\Big(\frac{\alpha_s C_A}{\pi n}\Big)^3-\frac{\pi^4}{120}\Big(\frac{\alpha_s C_A}{\pi n}\Big)^4+\frac{22}{5}\zeta_5\Big(\frac{\alpha_s C_A}{\pi n}\Big)^5+\Big(\frac{209}{9}\zeta_3^2-\frac{\pi^6}{1134}\Big)\Big(\frac{\alpha_s C_A}{\pi n}\Big)^6+...\,.
\end{align} }
To this leading logarithmic accuracy, we need only take $\beta(\alpha,\epsilon)=-\epsilon\alpha+...\,$ for the beta function. $\gamma^S$ satisfies the well-known result:
\begin{align}
  1&=\frac{\alpha_s C_A}{\pi n}\chi\Big(\gamma^S\Big)\,,\\
  \chi(\gamma)&=-2\gamma_E-\psi(\gamma)-\psi(1-\gamma)\,.
\end{align}
With a bit more effort, one can derive the all-orders form (valid to leading logarithmic accuracy) of $R^S$, see ref. \cite{Catani:1994sq}.

The appearance of the $R^{S}$ is driven by two causes, as explained in ref. \cite{Ciafaloni:2005cg}. First we can consider the entire splitting function as determined by the BFKL equation in $4-2\epsilon$ dimensions to exponentiate, including higher order terms in $\epsilon$. The second effect is a bit more subtle. Beyond the solution obtained from iterating the initial conditions consistent with perturbation theory (e.g. eq. \eqref{eq:perturbation_theory}), a general solution of the BFKL equation can be written as a ``sum'' over the power-law ansatz:
\begin{align}
\bar{\mathcal{F}}(n,\vec{k}_\perp)=\int\displaylimits_{c-i\infty}^{c+i\infty}\frac{d\gamma}{2\pi i}\Big(\vec{k}_\perp^{\,2}\Big)^{\gamma}f_{\epsilon}(n,\gamma)\,.
\end{align}  
Where $f_{\epsilon}(n,\gamma)$ must satisfy conditions obtained from plugging this ansatz into eq. \eqref{eq:BFKL_moment}. Recovering the BFKL solution which resums the PDF (that is, the solution which governs the structure twist-two operators) requires a saddle point evaluation of the $\gamma$-integral. If we write the $4-2\epsilon$ dimensional anomalous dimension in the form:
\begin{align}\label{eq:d_dim_Space_DGLAP}
\gamma^u(\alpha,n,\epsilon)&=\gamma^u(\alpha,n,0)+\epsilon\frac{\partial}{\partial \epsilon}\gamma^u(\alpha,n,0)+\frac{\epsilon^2}{2}\frac{\partial^2}{\partial \epsilon^2}\gamma^u(\alpha,n,0)+...
\end{align}
The end result is the conclusion (ref. \cite{Ciafaloni:2005cg}):
{\small\begin{align}\label{eq:space_coef_resum_ingredients}
  R^{S}\Big(\alpha_s,n\Big)&=\mathcal{N}^S\Big(\alpha_s,n\Big)\mathcal{R}^{S}\Big(\alpha_s,n\Big)\\
 \mathcal{R}^{S}\Big(\alpha_s,n\Big)&= \text{exp}\Bigg(-\int\displaylimits_{0}^{\alpha_s}\frac{d\alpha}{\alpha}\Big(\frac{\partial}{\partial \epsilon}\gamma^u(\alpha,n,0)-\frac{\beta(\alpha)}{2\alpha}\frac{\partial^2}{\partial \epsilon^2}\gamma^u(\alpha,n,0)+...\Big)\Bigg)
\end{align}  }
The function $\mathcal{N}$ is the so-called ``fluctuation factor'' resulting from the saddle point evaluation, and we have used the fact that the beta function in $4-2\epsilon$ dimensions is given as $\beta(\alpha,\epsilon)=-\epsilon\alpha-\beta(\alpha)$. $\mathcal{R}^{S}$ is the resummation of the coefficient function that is tied directly to the $\epsilon$-expansion of the DGLAP anomalous dimension. The determination of the fluctuation factor follows from the BFKL equation in $4-2\epsilon$ dimensions, but is not directly related to the $\epsilon$-expansion of the space-like DGLAP kernels. 

To all orders we would write the impact factor as: 
\begin{align}
  \int d^{2}\vec{k}_\perp\Theta(Q^2-\vec{k}_\perp^{\,2})\mathcal{F}\Big(n,\vec{k}_\perp\Big)&=\mathcal{N}^S\Big(\alpha_s,n\Big)\text{exp}\Big(\int\displaylimits_{0}^{A_{\epsilon}}\frac{d\alpha}{\beta(\alpha,\epsilon)}\gamma^u(\alpha,n,\epsilon)\Big)\,,\\
  A_{\epsilon}&=\alpha_s\times (Q^2/\mu^2)^{\epsilon}\,.
\end{align}  
The $\gamma^u(\alpha,n,\epsilon)$ is now the splitting function in $4-2\epsilon$. The $\epsilon$-expansion is:
{\small\begin{align}
  \int\displaylimits_{0}^{A_{\epsilon}}d\alpha\frac{\gamma^u(\alpha,n,\epsilon)}{\beta(\alpha,\epsilon)}&=-\int\displaylimits_{0}^{A_{\epsilon}}\frac{d\alpha}{\epsilon\alpha}\Bigg(\sum_{\ell=0}^{\infty}\epsilon^{\ell}\Big(-\frac{\beta(\alpha)}{\alpha}\Big)^{\ell}\Bigg)\Bigg(\sum_{k=0}^{\infty}\epsilon^k\frac{\partial^{k}}{\partial \epsilon^k}\gamma^u(\alpha,n,0)\Bigg)\nonumber\\
  &=\Big(-\frac{1}{\epsilon}+\text{ln}(Q^2/\mu^2)\Big)\gamma^{S}(\alpha_s,n)+\text{ln} \frac{R^S}{\mathcal{N}^S}+...\,.
\end{align}  }
Where the $...$ denote higher order poles in $\epsilon$ or terms which vanish as $\epsilon\rightarrow 0$.

Lastly, we remark that the coefficient function for the collinear factorization is then resummed via the formula:
{\small\begin{align}
C^{S}\Big(n,Q^2,\mu^2\Big)=\int\frac{d^{2}\vec{k}_\perp}{2\pi} h^S\Big(n,Q^2;\vec{k}_T^{\,2}\Big)\gamma^S(\alpha_s,n) \frac{R^S(\alpha_s,n)}{\vec{k}^{\,2}_\perp}\Big(\frac{\vec{k}_{\perp}^{\,2}}{\mu^2}\Big)^{\gamma^S(\alpha_s,n)}\,.
\end{align} }

\section{BMS/BK Duality and the Resummation of The Time-Like Anomalous Dimension}\label{sec:BMS_BFKL}
Given the BMS/BFKL duality discussed in the introduction, we postulate that we may also exhibit a ``$k_T$-factorization'' of the fragmentation cross-section, which we call a ``celestial factorization,'' since now instead of the BFKL equation operating in the transverse plane, it will evolve eikonal lines located on the celestial sphere. The form we want is:
\begin{align}\label{eq:bare_time_like_BFKL_factorization}
xD\Big(x,Q^2\Big)&=\int\displaylimits_{c-i\infty}^{c+i\infty}\frac{dn}{2\pi i} x^{-n}\int d^{2}\Omega_{\hat{b}}h^T\Big(n,\frac{\mu^2}{Q^2},n_a\cdot n_b,\Big)\bar{\mathcal{D}}\Big(n,\frac{\mu^2}{Q^2},n_a\cdot n_b\Big)\,,
\end{align}
The function $\mathcal{D}$ will satisfy the celestial BFKL equation in $4-2\epsilon$ dimensions. Written as an evolution equation in the energy fraction $x$, at lowest order in perturbation theory, this is simply:
{\small\begin{align}\label{eq:BFKL_for_Frag_One_Loop}
    x\frac{d}{dx}&\mathcal{D}\big(x,n_a\cdot n_b\big)=\SignOnPosBFKL(1+ 2\epsilon) \mathcal{D}\big(x,n_a\cdot n_b\big)\nonumber\\
    &\qquad\SignOnPosBFKL\Big(\frac{\mu e^{\frac{\gamma_E}{2}}}{xQ}\Big)^{2\epsilon}\frac{\alpha_sC_A}{\pi}\int\frac{d^{2-2\epsilon}\Omega_{\hat{j}}}{4\pi^{1-\epsilon}}\frac{n_a\!\cdot\! n_b}{n_a\!\cdot\! n_j\,n_j\!\cdot\! n_b}\Big(\mathcal{D}\big(x,n_a\!\cdot\! n_j\big)+\mathcal{D}\big(x,n_b\!\cdot\! n_j\big)-\mathcal{D}\big(x,n_a\!\cdot\! n_b\big)\Big)\,,
  \end{align}}
with the null direction $n_{j}=(1,\hat{j})$. An intuitive interpretation, at leading logarithmic accuracy in the large $N_c$ limit, is that $\mathcal{D}\big(x,n_a\cdot n_b\big)$ is the number of emissions with energy at least $xQ$ contained in the color singlet dipole with eikonal lines having the same opening angle as directions $n_a$ and $n_b$. We give a brief derivation of the equation in this limit in App. \ref{sec:derivation}, following the arguments of ref. \cite{Banfi:2002hw}. If we denote the loop expansion of $\mathcal{D}$ as:
\begin{align}
\mathcal{D}\big(x,n_a\cdot n_b\big)&=\sum_{i=0}^{\infty}\Big(\frac{\alpha_sC_A}{\pi}\Big)^{i+1}\mathcal{D}^{(i)}\big(x,n_a\cdot n_b\big)\,,
\end{align}
then this equation is to be solved with the boundary condition:
\begin{align}\label{eq:boundary_condition_time_like_BFKL}
\mathcal{D}^{(0)}\big(x,n_a\cdot n_b\big)&=\frac{1}{x^{1+2\epsilon}}\,.
\end{align}
Note that the boundary condition of eq. \eqref{eq:boundary_condition_time_like_BFKL} does not allow us to send $\epsilon\rightarrow 0$ in eq. \eqref{eq:BFKL_for_Frag_One_Loop}, and so eq. \eqref{eq:bare_time_like_BFKL_factorization} must be thought of as a \emph{bare} factorization of the fragmentation form factor. The initial condition given by perturbation theory is not a well-defined eigenfunction of the BFKL equation in 2 dimensions, having an infinite eigenvalue. This requires examining a formally regulated BFKL equation, even though naively collinear/ultraviolet divergences cancel in the equation.

Since we need the distribution of dipoles in the limit $n_a\cdot n_b\rightarrow 0$, that is, the collinear limit, we can expand eq. \eqref{eq:BFKL_for_Frag_One_Loop} in the collinear limit, which directly maps the time-like BFKL equation to the transverse plane, see ref. \cite{Marchesini:2003nh} as well as ref. \cite{Marchesini:2004ne}. Such an expansion is equivalent to projecting the celestial sphere to the tangent plane at the direction $n_a$. This side-steps the need to use the full conformal mapping between the effective theory of celestial eikonal lines embodied in the BMS equation and the infinite forward scattering eikonal line of the BK equation of ref. \cite{Hatta:2008st}. Thus we expect any resummation of the time-like anomalous dimension implied in eq. \eqref{eq:BFKL_for_Frag_One_Loop} to hold beyond strictly conformal theories, though differences may arise between the higher order corrections to the celestial and transverse plane BFKL equations.

When we project the celestial sphere to the tangent plane, we denote the transverse direction that celestial directions $\hat{a}$, $\hat{b}$, and $\hat{j}$ map to as $\vec{\theta}_{a}$,$\vec{\theta}_{b}$, and $\vec{\theta}_{j}$ respectively. We also introduce the shorthand:
\begin{align}
  \vec{\theta}_{ab}&=\vec{\theta}_a-\vec{\theta}_b\,,\nonumber\\
  \vec{\theta}_{aj}&=\vec{\theta}_a-\vec{\theta}_j\,,\nonumber\\
  \vec{\theta}_{jb}&=\vec{\theta}_j-\vec{\theta}_b\,.
\end{align}  
Performing this expansion using the results of App. \ref{sec:stereographics}, we have as our evolution equation:
{\small\begin{align}
\mathcal{D}\big(x,n_a\cdot n_b\big)&\rightarrow\mathcal{D}\big(x,\vec{\theta}_{ab}^{\,2}\big),\,\\
\label{eq:BFKL_evolution_small_angle}x\frac{d}{dx}\mathcal{D}\big(x,\vec{\theta}_{ab}^{\,2}\big)&=\SignOnPosBFKL (1+2\epsilon) \mathcal{D}\big(x,\vec{\theta}_{ab}^{\,2}\big)\nonumber\\
&\quad\SignOnPosBFKL\Big(\frac{\mu e^{\frac{\gamma_E}{2}}}{xQ}\Big)^{2\epsilon}\frac{\alpha_sC_A}{\pi}\int\frac{d^{2-2\epsilon}\vec{\theta}_{j}}{2\pi^{1-\epsilon}}\frac{\vec{\theta}_{ab}^{\,2}}{\vec{\theta}_{aj}^{\,2}\vec{\theta}_{jb}^{\,2}}\Big(\mathcal{D}\big(x,\vec{\theta}_{aj}^{\,2}\big)+\mathcal{D}\big(x,\vec{\theta}_{jb}^{\,2}\big)-\mathcal{D}\big(x,\vec{\theta}_{ab}^{\,2}\big)\Big)\,.
\end{align}  }
And to the same accuracy we have:
\begin{align}\label{eq:bare_time_like_BFKL_factorization_small_angle}
xD\Big(x,\frac{\mu^2}{Q^2}\Big)&=\int\displaylimits_{c-i\infty}^{c+i\infty}\frac{dn}{2\pi i} x^{-n}\int \frac{d^{2}\vec{\theta}}{2\pi}\Theta\Big(1-\vec{\theta}^{\,2}\Big)h^T\Big(\vec{\theta}^{\,2},n,\frac{\mu^2}{Q^2}\Big)\bar{\mathcal{D}}\Big(n,\vec{\theta}^{\,2};\frac{\mu^2}{Q^2}\Big)\,.
\end{align}

\subsection{Celestial BFKL in $4-2\epsilon$ Dimensions}
We now do the same analysis as in Sec. \ref{sec:space_like_BFKL} for eq. \eqref{eq:BFKL_for_Frag_One_Loop}. When we perform the moment space transform, we have for the derivative term:
\begin{align}
\int_0^{1}dx x^n\Big(x\frac{d}{dx}\mathcal{D}\big(x,\vec{\theta}_{ab}^{\,2}\big)\Big)=\mathcal{D}\Big(1,\vec{\theta}_{ab}^{\,2}\Big)-(1+n)\bar{\mathcal{D}}\Big(n,\vec{\theta}_{ab}^{\,2}\Big)\,.
\end{align}  
We identify $\mathcal{D}\Big(1,\vec{\theta}_{ab}^{\,2}\Big)$ as the appropriate boundary condition in the moment space evolution equation, so we have:
{\small\begin{align}
  &\bar{\mathcal{D}}\Big(n,\vec{\theta}_{ab}^{\,2}\Big) =\frac{d\Big(\vec{\theta}_{ab}^{\,2}\Big)}{n-2\epsilon}+\frac{\alpha_sC_A}{\pi (n-2\epsilon)}K\otimes \bar{\mathcal{D}}\Big(n-2\epsilon,\vec{\theta}_{ab}^{\,2}\Big)\label{eq:BFKL_for_Frag_One_Loop_moment},\\
  K\otimes &\bar{\mathcal{D}}\Big(n-2\epsilon,\vec{\theta}_{ab}^{\,2}\Big)\nonumber\\
  &\quad=e^{\epsilon\gamma_E}\Big(\frac{\mu^2e^{\frac{\gamma_E}{2}}}{Q^2}\Big)^{\epsilon}\frac{\alpha_sC_A}{\pi}\int\frac{d^{2-2\epsilon}\vec{\theta}_{j}}{2\pi^{1-\epsilon}}\frac{\vec{\theta}_{ab}^{\,2}}{\vec{\theta}_{aj}^{\,2}\vec{\theta}_{jb}^{\,2}}\Big(\bar{\mathcal{D}}\big(n-2\epsilon,\vec{\theta}_{aj}^{\,2}\big)+\bar{\mathcal{D}}\big(n-2\epsilon,\vec{\theta}_{jb}^{\,2}\big)-\bar{\mathcal{D}}\big(n-2\epsilon,\vec{\theta}_{ab}^{\,2}\big)\Big)\,,\nonumber
\end{align}  }
Since the action on a power-law ansatz produces the same result, we denote the action of both the position space and momentum space BFKL with the same notation, $K\otimes g$ for test function $g$. We wish to use a similar power-law initial condition, so defining:
{\small\begin{align}
  d\Big(\vec{\theta}_{ab}^{\,2}\Big)&=\frac{\alpha_sC_A}{\pi}\Big(\frac{Q^2 \vec{\theta}_{ab}^{\,2}}{\mu^2}\Big)^{\delta}\,,\\
  K\otimes\Big(\Big(\frac{Q^2 \vec{\theta}_{ab}^{\,2}}{\mu^2}\Big)^{\delta}\Big)&=I(\delta,\epsilon)\Big(\frac{Q^2 \vec{\theta}_{ab}^{\,2}}{\mu^2}\Big)^{\delta-\epsilon}\,.
\end{align}}  
The function $I$ is identical to eq. \eqref{eq:I_def}. The result for the iterative expansion is:
{\small\begin{align}
  \bar{\mathcal{D}}\Big(n,\vec{\theta}_{ab}^{\,2}\Big)&=\Big(\frac{Q^2 \vec{\theta}_{ab}^{\,2}}{\mu^2}\Big)^{\delta}\sum_{\ell=0}^{\infty}c_{\ell}(\delta,\epsilon)\Big[\frac{\alpha_sC_A}{\pi}\Big(\frac{\mu^2}{Q^2\vec{\theta}_{ab}^{\,2}}\Big)^{\epsilon}\Big]^{\ell}\prod_{i=0}^{\ell}(n-2\epsilon(i+1))^{-1}\,.
\end{align}}
The $c$'s are defined identically to eq. \eqref{eq:c_def}. We now write:
\begin{align}
\prod_{i=0}^{\ell}(n-2\epsilon (i+1))^{-1}&=(2\epsilon)^{-\ell}\frac{\Gamma\Big(\frac{n}{2\epsilon}-\ell\Big)}{\Gamma\Big(\frac{n}{2\epsilon}\Big)}=\frac{1}{(2\epsilon)^{\ell} \Big(\frac{n}{2\epsilon}-1\Big)_{\ell}}
\end{align}  
The notation $(x)_k$ denotes the so-called falling Pochhammer symbol. 

Again setting $\delta=0$, then in the limit that $\epsilon\rightarrow 0$, we find the result:
\begin{align}\label{eq:celestial_BFKL_soln}
\bar{\mathcal{D}}\Big(n,\vec{\theta}_{ab}^{\,2}\Big)&=\gamma^T(\alpha_s,n)R^T(\alpha_s,n)\Big(\frac{Q^2\vec{\theta}_{ab}^{\,2}}{\mu^2}\Big)^{\gamma^T(\alpha_s,n)}\text{exp}\Bigg(\int\displaylimits_0^{\alpha_s(\mu^2)}\frac{d\alpha}{\beta(\alpha,\epsilon)}\gamma^T(\alpha,n)\Bigg)\,.
\end{align} 
Then to match the pure glue process $\phi\rightarrow h+X$, calculated via eq. \eqref{eq:bare_time_like_BFKL_factorization_small_angle} to least to LL accuracy, we would find:
\begin{align}
h^{T}\Big(\vec{\theta}_{ab},n,\frac{\mu^2}{Q^2}\Big)=\frac{1}{\vec{\theta}_{ab}^{\,2}}\,.
\end{align}  
Now it is instructive to calculate the resummation predicted from eq. \eqref{eq:BFKL_for_Frag_One_Loop_moment}. We find to seven loops in the anomalous dimension that:
{\small\begin{align}
  \gamma^T\Big|_{\text{eq.} \eqref{eq:BFKL_for_Frag_One_Loop_moment}}&=\frac{\alpha_s C_A}{\pi n}\Bigg(1-2\Big(\frac{\alpha_s C_A}{\pi n^2}\Big)+8\Big(\frac{\alpha_s C_A}{\pi n^2}\Big)^2-40\Big(\frac{\alpha_s C_A}{\pi n^2}\Big)^3\nonumber\\
  &\qquad\qquad+224\Big(\frac{\alpha_s C_A}{\pi n^2}\Big)^4-1334\Big(\frac{\alpha_s C_A}{\pi n^2}\Big)^5+8448\Big(\frac{\alpha_s C_A}{\pi n^2}\Big)^6+...\Bigg)\nonumber\\
  &\qquad+2\zeta_3\Big(\frac{\alpha_s C_A}{\pi n}\Big)^4\Bigg(1-10\Big(\frac{\alpha_s C_A}{\pi n^2}\Big)+84\Big(\frac{\alpha_s C_A}{\pi n^2}\Big)^2-672\Big(\frac{\alpha_s C_A}{\pi n^2}\Big)^3+...\Bigg)\nonumber\\
  &\qquad+2\zeta_5\Big(\frac{\alpha_s C_A}{\pi n}\Big)^6\Bigg(1-14\Big(\frac{\alpha_s C_A}{\pi n^2}\Big)+...\Bigg)\nonumber\\
  &\qquad+12\zeta_3^2\Big(\frac{\alpha_s C_A}{\pi n}\Big)^7+...
\end{align} }
This precisely agrees with the result for the leading log anomalous dimension (given on the first line), but includes subleading contributions as well. Indeed, a feature of the celestial BFKL approach to resummation of the soft region of the time-like DGLAP evolution is that it fails to respect a manifest log counting described in eq. \eqref{eq:log_count_time}. The leading order celestial BFKL equation not only resums the LL anomalous dimension and coefficient function, but also terms at all sub-leading logarithmic orders.  Thus it will produce partial results at each subleading logarithmic order. So instead we will introduce a form of angular-ordered DGLAP evolution (eq. \eqref{eq:generalized_resummation_equation_for_DGLAP_momentum_space}) which will accomplish the same resummation, but with manifest logarithmic accuracy.

However, though it does not have manifest log counting, when we compare the terms in the anomalous dimension that the celestial BFKL equation does resum with either known results or with the angular-ordered DGLAP evolution we introduce below, we find complete agreement.\footnote{Specifically, we can check the tower of logarithms associated with the terms of the form $p_{i,-i-1,0}$, in the notation of Sec. \ref{sec:resum_time}. We have explicitly checked that up to 9 loops, the two approaches agree.} As will be explained in detail in Sec. \ref{sec:conjectures}, each term in the laurent expansion of the space-like anomalous dimension acts as a seed for a tower of terms that must be resummed in the time-like anomalous dimension. The celestial BFKL equation has access to only those seeds in the space-like anomalous dimension predicted from the leading order transverse-plane BFKL equation, those that are LL in the space-like counting. It correctly resums the tower of terms associated to those seeds. 

Beyond the anomalous dimension, the celestial BFKL equation also correctly reproduces the leading log resummation of the coefficient function $R^T$. However when we compare the resummation of the coefficient function implied by the angular-ordered DGLAP evolution to the one resulting from the celestial BFKL, we will see a discrepancy starting at three-loop order in the $\NNNLL$ resummation (in the time-like log counting) in the expansion of $R^T$. This is precisely when the fluctuation factor of the space-like case has its first non-trivial contribution. We will discuss this in more detail in Sec. \ref{celestial_BFKL_versus_angular_DGLAP}. But first we introduce the angular-ordered DGLAP evolution in $4-2\epsilon$ dimensions, make an all-orders conjecture about its kernel, and show how it reproduces known results concerning the resummation of the time-like anomalous dimension to $\NNNLL$ order, and the coefficient function to $\NNLL$ order. 

\section{Angular Regularization of the Soft Region of Fragmentation and Reciprocity}\label{sec:conjectures}
We will posit a factorization for the soft region of the fragmentation cross-section, a factorization which we conjecture will resum all pole terms in the laurent expansion about $n=0$ in the time-like anomalous dimension, while leaving us a puzzle about the coefficient function. The factorization will imply an angular-ordered evolution equation, which we write as valid to all-orders in dimensional regularization. We write:
\begin{align}\label{eq:factor_angular_frag_func}
xD(x,R^2,Q^2)&=\int_{x}^{1}\frac{dz}{z}\frac{x}{z}d\Big(\frac{x}{z},R^2,R_f^2,\mu^2,Q^2\Big)z\tilde{C}^{T}\Big(z,R_f^2,\mu^2,Q^2\Big)\,.
\end{align}  
We have replaced the infra-red scale $\Lambda$ in eq. \eqref{eq:basic_factorization_FF} with an angular cutoff $R$. This angular cutoff is understood to be the minimum angle that any other parton can approach the fragmented particle carrying energy fraction $x$. $R_f$ is the factorization angle, where we separate out the collinear dynamics from the large angle ultra-violet processes. We are interested in the limit $R\rightarrow 0$, where we recover the form factor $xD(x,Q^2)$ in dimensional regularization, eq. \eqref{eq:Factor_IR_Divergences}. In what follows, when we take $R_f=1$ and $\mu=Q$, writing:
\begin{align}
d\Big(x,R^2,R_f^2,\mu^2,Q^2\Big)\Bigg|_{R_f=1}&= d\Big(x,R^2,\mu^2,Q^2\Big),\\
d\Big(x,R^2,R_f^2,\mu^2,Q^2\Big)\Bigg|_{\mu=Q,R_f=1}&= d\Big(x,R^2\Big).
\end{align}
Note however that $\tilde{C}^{T}$ is not the same as the matching coefficient found in eq. \eqref{eq:Factor_IR_Divergences}. In the limit $R\rightarrow 0$, the function $d$ will resum the $x\rightarrow 0$ behavior of the coefficient function and time-like DGLAP anomalous dimension, so that we write for the mellin transformed function (up to the puzzle at $\NNNLL$ order we discuss in Sec. \ref{celestial_BFKL_versus_angular_DGLAP}):
\begin{align}\label{eq:define_coef_func_Angular_Order}
\lim_{R\rightarrow 0}\bar{d}\Big(n,R^2\Big)&=\mathcal{R}^{T}(\alpha_s,n)\text{exp}\Big(\int_{0}^{\alpha_s}\frac{d\alpha}{\beta(\alpha,\epsilon)}\gamma^T(\alpha,n)\Big)\,.
\end{align}  
After solving the angular-ordered evolution equation, we will have resummed the time-like coefficient function with the factor $\mathcal{R}^{T}$, and the anomalous dimension $\gamma^T$. Moreover, both will be determined by the DGLAP anomalous dimension $\gamma^u(\alpha,n,\epsilon)$ in $4-2\epsilon$ dimensions described in Sec. \ref{sec:space_like_BFKL}, and the resulting $\gamma^T$ will enjoy the reciprocity relation with the space-like anomalous dimension:
\begin{align}\label{eq:reciprocity}
\gamma^{S}\Big(n+2\gamma^T(n)\Big)=\gamma^T(n)\,.
\end{align}  
In App. \ref{sec:reciprocity}, we describe how resummation of the time-like anomalous dimension follows from this reciprocity relation in a pure Yang-Mills theory.

\subsection{Angular Evolution}
\begin{figure}\center
  \includegraphics[width=0.55\textwidth]{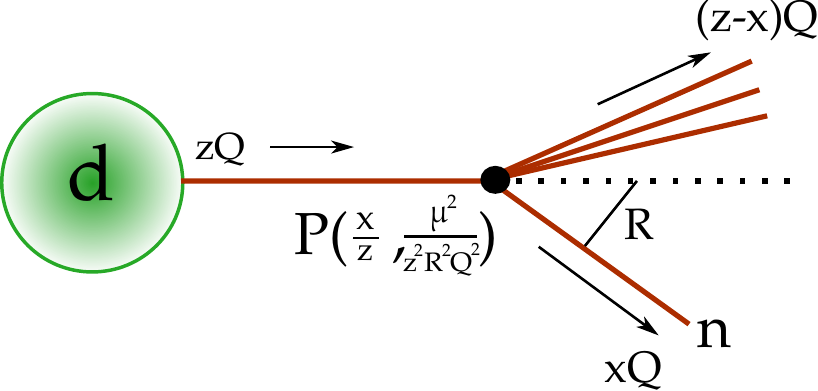}
  \caption{\label{fig:angular_order}The kinematic structure of the angular-ordered evolution. Note that the direction of the momentum of the fragmented hadron carrying energy $xQ$ defines the light-cone direction $n$. Therefore, the $\mu$ depedence of the splitting kernel $P$ tracks the transverse momentum of the parent parton with respect to the null direction $n$.}
\end{figure}  
 The angular evolution equation within dimensional regularization which resums the $x\rightarrow 0$ limit of the fragmentation form factor will have the form: 
{\small\begin{align}\label{eq:generalized_resummation_equation_for_DGLAP_momentum_space}
  R^2\frac{\partial}{\partial R^2} x^{1+2\epsilon} d\Big(x,R^2,\mu^2,Q^2\Big)&=\rho\Big(\frac{\mu^2}{R^2Q^2}\Big)x^{1+2\epsilon} d\Big(x,R^2,\mu^2,Q^2\Big)\nonumber\\
  &\qquad+\int_{x}^{1}\frac{dz}{z}P\Big(\frac{x}{z};\frac{\mu^2}{z^2R^2Q^2}\Big)z^{1+2\epsilon} d\Big(z,R^2,\mu^2,Q^2\Big)\,,\\
  P\Big(\frac{x}{z};\frac{\mu^2}{z^2R^2Q^2}\Big)&=\sum_{\ell=1}^{\infty} P^{(\ell-1)}\Big(\frac{x}{z};a_s;\epsilon\Big)\,\Big(\frac{\mu^2}{z^2R^2Q^2}\Big)^{\ell\epsilon}\,,\label{eq:expansion_of_kernel_P_momentum_space}\\
  \rho\Big(\frac{\mu^2}{R^2Q^2}\Big)&=\sum_{\ell=1}^{\infty}\rho^{(\ell-1)}(a_s;\epsilon)\Big(\frac{\mu^2}{R^2Q^2}\Big)^{\ell\epsilon}\,.
\end{align}}
$P$ is the splitting kernel which depends on the kinematics of the splitting process. We illustrate the kinematic structure of the evolution equation in Fig. \ref{fig:angular_order}. We note that the splitting kernel depends upon the transverse momentum of the parent ($zRQ$) with respect to the direction of the fragmented hadron. The fragmented hadron defines the light-cone direction $n$ of the fragmentation function. We have also expanded the anomalous dimension in eq. \eqref{eq:expansion_of_kernel_P_momentum_space} according to the scaling in dimensional regularization parameter $\mu$. If the coupling constant did not run, $\beta(\alpha_s)=0$, this expansion would correspond exactly to the loop expansion. However, as will be explicitly shown below, counterterms and divergences arising from the renormalization of the coupling constant interfere with a strict identification of the scaling in dimensional regularization and the order of the loop expansion.

To actually calculate the small-x logarithms in the coefficient function and the anomalous dimension, we take the moment of both sides of eq. \eqref{eq:generalized_resummation_equation_for_DGLAP_momentum_space} after multiplying by $x^{-2\epsilon}$ to derive the moment-space formulation:
{\small\begin{align}\label{eq:generalized_resummation_equation_for_DGLAP_moment_space}
  R^2\frac{\partial}{\partial R^2} \bar d(n,R^2,\mu^2,Q^2)&=\rho\Big(\frac{\mu^2}{R^2Q^2}\Big)\bar d(n,R^2,\mu^2,Q^2)\nonumber\\
  &\qquad+\sum_{\ell=1}^{\infty}\bar P^{(\ell-1)}\Big(n-2\epsilon;a_s;\epsilon\Big)\Big(\frac{\mu^2}{R^2Q^2}\Big)^{\ell\epsilon}\bar d\Big(n-2\ell\epsilon,R^2,\mu^2,Q^2\Big)\,.
\end{align}  }
Further, we introduce the function:
{\small\begin{align}
\bar{P}\Big(n,\epsilon,\frac{\mu^2}{R^2Q^2}\Big)=\sum_{\ell=1}^{\infty}\bar{P}^{(\ell-1)}(n,\epsilon)\Big(\frac{\mu^2}{R^2Q^2}\Big)^{\ell\epsilon}=\int_{0}^{1}\frac{dx}{x}x^{n}\Bigg(x\sum_{\ell=1}^{\infty}P^{(\ell-1)}(x,\epsilon)\Big(\frac{\mu^2}{R^2Q^2}\Big)^{\ell\epsilon}\Bigg)\,,
\end{align}}
 The key claim will be that instead of using the celestial BFKL, after solving for $d$ using the above equation, we can reconstruct $\mathcal{D}$ of the factorization in eq. \eqref{eq:bare_time_like_BFKL_factorization_small_angle} via:\footnote{Caveats to this equation will be discussed in Sec. \ref{celestial_BFKL_versus_angular_DGLAP}}
\begin{align}\label{eq:celestial_BFKL_to_Frag}
\lim_{R\rightarrow 0}\bar{d}\Big(n,R^2,\frac{\mu^2}{Q^2}\Big)&\sim\int \frac{d^2\vec{\theta}}{2\pi\vec{\theta}^{\,2}}\Theta\Big(1-\vec{\theta}^{\,2}\Big)\bar{\mathcal{D}}\Big(n,\vec{\theta}^{\,2};\frac{\mu^2}{Q^2}\Big)
\end{align}  
We can then define the precise relationship between the kernel $P$ of the angular-ordered DGLAP evolution within dimensional regularization and \emph{traditional} transverse-plane BFKL factorization found in eq. \eqref{eq:impact_factor_collinear_factorization}:
\begin{align}\label{eq:fixing_angular_ordering_kernel}
  \int_0^{1}\frac{dR^2}{R^2}\Big(\rho(R^{-2})+\bar{P}(n,\epsilon,R^{-2})\Big)&=\int_{0}^{\alpha_s}\frac{d\alpha}{\beta(\alpha,\epsilon)}\gamma^u(\alpha,n,\epsilon).
\end{align}
Here $\gamma^u$ is the $4-2\epsilon$ dimensional DGLAP anomalous dimension determined by the BFKL equation in $4-2\epsilon$-dimensions introduced in eq. \eqref{eq:d_dim_Space_DGLAP}. We note that eq. \eqref{eq:fixing_angular_ordering_kernel} is to hold  order-by-order in the $\epsilon$ and loop-expansions, \emph{including the finite terms as $\epsilon\rightarrow 0$}.

Additionally, we will find that to reproduce known results for the resummation of the coefficient function and to connect to resummation of the anomalous dimension resulting from the reciprocity relation of refs. \cite{Dokshitzer:2005bf,Basso:2006nk}, we find that we would identify:
\begin{align}\label{eq:determine_rho}
\rho(1) &= \frac{\beta(\alpha_s)}{\alpha_s}\,.
\end{align}
Thus we see that the $\rho$ term in the angular ordered evolution equation acts as a type of conformal anomaly to the angular-ordered evolution. This is not unexpected from the map between traditional and celestial BFKL: we know this map strictly holds in 4 dimensions for a conformal theory. One would then expect at higher orders, conformal symmetry breaking terms need to be accounted for.

\subsection{Structure of the Angular-Ordered Evolution Kernel}\label{sec:resum_time}

Since we work to all orders in $\epsilon$, we note that the renormalization of ultra-violet divergences of the theory associated to $\alpha_s$ requires the following expansion for the anomalous dimension:
{\small\begin{align}
  a_s&=\frac{\alpha_sC_A}{\pi}\\
\label{eq:kernel_1}  \bar P^{(0)}(n,a_s,\epsilon)&=a_s \bar p_{0}(n,\epsilon)\Big(1-a_s\frac{\beta_0}{\epsilon}-a_s^2\frac{\beta_1}{2\epsilon}+a_s^2\frac{\beta_0^2}{\epsilon^2}-a_s^3\frac{\beta_0^3}{\epsilon^3}+...\Big)\\
\bar P^{(1)}(n,a_s,\epsilon)&=a_s\bar p_{0}(n,\epsilon)\Big(a_s\frac{\beta_0}{\epsilon}-a_s^2\frac{2\beta_0^2}{\epsilon^2}+a_s^3\frac{3\beta_0^3}{\epsilon^3}+...\Big)\nonumber\\
&\qquad+a_s^2 \Big(1-a_s\frac{\beta_0}{\epsilon}-a_s^2\frac{\beta_1}{2\epsilon}+a_s^2\frac{\beta_0^2}{\epsilon^2}-a_s^3\frac{\beta_0^3}{\epsilon^3}+...\Big)^2\bar p_{1}(n,\epsilon)\\
  \bar P^{(2)}(n,a_s,\epsilon)&=a_s\bar p_{0}(n,\epsilon)\Big(a_s^2\frac{\beta_0^2}{\epsilon^2}-a_s^3\frac{3\beta_0^3}{\epsilon^3}+a_s^2\frac{\beta_1}{2\epsilon}+...\Big)\nonumber\\
&\qquad+a_s^3 \Big(1-a_s\frac{\beta_0}{\epsilon}-a_s^2\frac{\beta_1}{2\epsilon}+a_s^2\frac{\beta_0^2}{\epsilon^2}-a_s^3\frac{\beta_0^3}{\epsilon^3}+...\Big)^3\bar p_{2}(n,\epsilon)+...\\
  \label{eq:kernel_4}  \bar P^{(3)}(n,a_s,\epsilon)&=a_s\bar p_{0}(n,\epsilon)\Big(a_s^3\frac{\beta_0^3}{\epsilon^3}+...\Big)\nonumber\\
&\qquad+a_s^4 \Big(1-a_s\frac{\beta_0}{\epsilon}-a_s^2\frac{\beta_1}{2\epsilon}+a_s^2\frac{\beta_0^2}{\epsilon^2}-a_s^3\frac{\beta_0^3}{\epsilon^3}+...\Big)^3\bar p_{3}(n,\epsilon)+...,
\end{align}  }
and so on to higher orders. Here we have included all terms that will be necessary to $\NNNLL$ accuracy. The functions $\bar{p}_{i}$ themselves have a simultaneous expansion in $n$ and $\epsilon$:
\begin{align}
\bar{p}_{i}(n,\epsilon)&=\sum_{j=-i-1}^{\infty}\sum_{k=0}^{\infty}n^{j}\epsilon^{k}p_{i,j,k}\,.
\end{align}  
The single logarithmic structure of the kernel $P$ implies that the highest pole in $n$ for $\bar p_{i}(n,\epsilon)$ is of order $i+1$. Since we take the leading-logarithmic anomalous dimension to be an order $O(n)$ quantity, then we must adopt the power counting that $a_s\sim n^2$, so that we will find that to N$^3$LL accuracy in the anomalous dimension and $\NNLL$ in the coefficient, we will need the terms:
{\small\begin{align}
    \bar p_0(n)&=\Big(\frac{1}{n}\Big(p_{0,-1,0}+\epsilon p_{0,-1,1}+\epsilon^2 p_{0,-1,2}\Big)+p_{0,0,0}+\epsilon p_{0,0,1}+\epsilon^2 p_{0,0,2}\nonumber\\
    &\qquad\qquad\qquad+n (p_{0,1,0}+\epsilon p_{0,1,1})+n^2p_{0,2,0}+...\Big)\,,\\
\bar p_1(n)&=\frac{1}{n^{2}}\Big(p_{1,-2,0}+\epsilon p_{1,-2,1}\Big)+\frac{1}{n}\Big(p_{1,-1,0}+\epsilon p_{1,-1,1}\Big)+p_{1,0,0}+..\,,\\
\bar p_2(n)&=\frac{1}{n^{3}}\Big(p_{2,-3,0}+\epsilon p_{2,-3,1}\Big)+\frac{1}{n^{2}}\Big(p_{2,-2,0}\Big)+..\,,\\
\bar p_3(n)&=\frac{p_{3,-4,0}}{n^{4}}+..\,.
\end{align}}
Using the relation of eq. \eqref{eq:fixing_angular_ordering_kernel}, we can map the $p_{i,j,k}$ to the appropriate values of the expansion of the space-like DGLAP anomalous dimension defined in eq. \eqref{eq:laurent_space}, using also eq. \eqref{eq:determine_rho}:
\begin{align}\label{eq:gamma_s_to_p_start}
\gamma^S_{0,-1}&=p_{0,-1,0}\,,\qquad\gamma^S_{0,0}=p_{0,0,0}+\beta_0\,,\qquad\gamma^S_{0,j}=p_{0,j,0}\text{ if }j>0\,,\\
\gamma^S_{1,-2}&=p_{1,-2,0}\,,\qquad\gamma^S_{1,-1}=p_{1,-1,0}-\beta_0p_{0,-1,1}\,,\qquad\gamma^S_{1,0}=p_{1,0,0}+\beta_1-\beta_0 p_{0,0,1}\,,\\
\gamma^S_{2,-3}&=p_{2,-3,0}\,,\qquad\gamma^S_{2,-2}=p_{2,-2,0}-\beta_0 p_{1,-2,1}\,,\\
\gamma^S_{3,-4}&=p_{3,-4,0}\,.\label{eq:gamma_s_to_p_end}
\end{align}  
We collect the actual values of $\gamma^{S}_{i,j}$ and  $p_{i,j,k}$ in App. \ref{sec:constants}.

\subsection{Solution}
The eq. \eqref{eq:generalized_resummation_equation_for_DGLAP_moment_space} is not straightforward to solve. However, we will find that with an appropriate ansatz for the solution in the ordered limits that $R\rightarrow 0$ and then $\epsilon\rightarrow 0$, the all-orders form of the resummed anomalous dimension and coefficient function can be deduced from a few terms in its iterative expansion. First we can exponentiate the $\rho$ factor from the eq. \eqref{eq:generalized_resummation_equation_for_DGLAP_moment_space}, so in moment space, the first few iterations for the function $\bar{d}$ are:
{\small\begin{align}
    \text{exp}\Bigg(-&\int_{R^2}^{1}\frac{d\theta^2}{\theta^2}\rho\big(\theta^{-2}\big) \Bigg)\bar{d}\Big(n,R^2\Big)\nonumber\\
    &=1+\sum_{\ell_1}\powR\int_{R^2}^{1}\frac{d\theta_1}{\theta_1^{1+2\ell_1\epsilon}}\bar{P}^{(\ell_1-1)}(n-2\epsilon)\nonumber\\
  &\qquad+\powR^2\sum_{\ell_1,\ell_2}\int_{R}^{1}\frac{d\theta_1}{\theta_1^{1+2\ell_1\epsilon}}\int_{\theta_1}^{1}\frac{d\theta_2}{\theta_2^{1+2\ell_2\epsilon}}\bar{P}^{(\ell_1-1)}(n-2\epsilon)\bar{P}^{(\ell_2-1)}\Big(n-2\epsilon(1+\ell_1)\Big)\nonumber\\
  &\qquad+\sum_{\ell_1,\ell_2,\ell_3}\powR^3\int_{R}^{1}\frac{d\theta_1}{\theta_1^{1+2\ell_1\epsilon}}\int_{\theta_1}^{1}\frac{d\theta_2}{\theta_2^{1+2\ell_2\epsilon}}\int_{\theta_2}^{1}\frac{d\theta_3}{\theta_3^{1+2\ell_3\epsilon}}\nonumber\\
  &\qquad\qquad\qquad\bar{P}^{(\ell_1-1)}(n-2\epsilon)\bar{P}^{(\ell_2-1)}\Big(n-2\epsilon(1+\ell_1)\Big)\bar{P}^{(\ell_3-1)}\Big(n-2\epsilon(1+\ell_1+\ell_2)\Big)+...
  \end{align}}
All sums are from $\ell_{i}=1$ to $\infty$. Recognizing the pattern, we can then introduce the functions:
{\small\begin{align}
  I\Big(\ell_1;n;R\Big)&=\powR\int_{R}^{1}\frac{d\theta_{1}}{\theta_1^{1+2\ell_1\epsilon}}\bar{P}^{(\ell_1-1)}(n-2\epsilon),\nonumber\\
  I\Big(\ell_1,...,\ell_k;n;R\Big)&=\powR^k\int_{R}^{1}\frac{d\theta_{1}}{\theta_1^{1+2\ell_1\epsilon}}\bar{P}^{(\ell_1-1)}(n-2\epsilon)\int_{R}^{1}\prod_{i=2}^{k}\frac{d\theta_{i}}{\theta_i^{1+2\ell_i\epsilon}}\Theta(\theta_{i}-\theta_{i-1})\bar{P}^{(\ell_i-1)}\Big(n-2\epsilon(1+\sum_{j=1}^{i-1}\ell_{j})\Big)\,,\text{if } k>1\,.
\end{align}}
Then the expansion is easily obtained to be:
{\small\begin{align}\label{eq:iterative_expansion_fin}
\bar{d}\Big(n,R^2\Big)&=\text{exp}\Bigg(\int_{R^2}^{1}\frac{d\theta^2}{\theta^2}\rho\big(\theta^{-2}\big) \Bigg)\Bigg(1+\sum_{\ell_1}I(\ell_1;n;R)+\sum_{\ell_1,\ell_2}I(\ell_1,\ell_2;n;R)+\sum_{\ell_1,\ell_2,\ell_3}I(\ell_1,\ell_2,\ell_3;n;R)+...\Bigg)\,\,.
\end{align}}
Again all sums are from $\ell_{i}=1$ to $\infty$. We evaluate the angular integrals as:
\begin{align}
\lim_{R\rightarrow 0}\int_{R}^{1}\frac{d\theta_1}{\theta_1}\theta_1^{-2\ell_{1}\epsilon}\int_{R}^{1}\prod_{j=2}^{k}\Theta(\theta_{j}-\theta_{j-1})\frac{d\theta_j}{\theta_j}\theta_j^{-2\ell_{j}\epsilon}=(2\epsilon)^{-k}\prod_{j=1}^{k}\Bigg(\sum_{i=1}^{j}\ell_{i}\Bigg)^{-1}.
\end{align}  
Then we factor the IR divergences into a renormalization factor $Z$ when $R=0$ as:
\begin{align}\label{eq:Factor_IR_Divergences_remind}
  \bar d\Big(n,0\Big)&=\text{exp}\Big(\int_{0}^{\alpha_s}\frac{d\alpha}{\beta(\alpha,\epsilon)}\gamma^T(\alpha,n)\Big)\,\mathcal{R}^T(\alpha_s,n).
\end{align}
Where $\mathcal{R}^T$ is finite as $\epsilon\rightarrow 0$, and $\gamma^T$ is the resummed component of the time-like anomalous dimension. To compute the anomalous dimension generated by the eq. \eqref{eq:generalized_resummation_equation_for_DGLAP_moment_space}, we take the log of the iterations, take the $\frac{1}{\epsilon}$ term in the limit $\epsilon\rightarrow 0$, and then take the derivative $\alpha_s\frac{\partial}{\partial\alpha_s}$. Therefore we note that once we write $\text{ln}\bar{d}$ as an expansion in $\epsilon$, we can extract the anomalous dimension as:
\begin{align}
  \text{ln}\bar{d}&=\sum_{k=-\infty}^{\infty}\epsilon^kg_{k}\\
  g_{-1}&=-\frac{1}{\epsilon}\int_{0}^{\alpha_s}\frac{d\alpha}{\alpha}\Big\{\gamma^T(\alpha,n)\Big\}\\
  \gamma^T&=-\alpha_s\frac{d}{d\alpha_s}g_{-1}
\end{align}
We then decompose the anomalous dimension into components which produce the poles as $n\rightarrow 0$, once expanded in $\alpha_s$ and all terms which are regular as $n\rightarrow 0$.  All pole terms are collected in the function $\gamma$:
{\small\begin{align}\label{eq:gamma_def}
    \gamma^T&=\frac{n}{2}\gamma+\sum_{i=0}^{\infty}\sum_{j=0}^{\infty} a_s^{i+1} n^j b_{i,j}+\rho(\alpha_s)
  \end{align}}
We then posit the ansatz that this $\gamma$ will satisfy an equation:
\begin{align}\label{eq:anom_dim_eq_resum}
  \gamma^2+\gamma&=\Gamma\frac{a_s}{n^2}\Bigg(a_{0,0}+a_{1,1} n\gamma +a_{1,0} n + a_{2,2} n^2\gamma^2+ a_{2,1} n^2\gamma+ a_{2,0} n^2\nonumber\\
  &\qquad+ a_{3,3}n^3\gamma^3+a_{3,2}n^3\gamma^2+a_{3,1}n^3\gamma+a_{3,0}n^3+...\Bigg)
\end{align}
We have written the expansion such that if we perform the logarithmic expansion of the anomalous dimension with the power counting that $\frac{a_s}{n^2}\sim O(1)$, then the coefficients $a_{k,\ell}$ enter at the N$^{k}$LL order. Further, eq. \eqref{eq:anom_dim_eq_resum} was written with explicit dependence on $a_s$ and $n$, i.e., the coefficients $a_{k,\ell}$ are understood to be pure numbers, not functions of $a_s$ and $n$. While $\gamma$ is the function that resums the anomalous dimension, and the coefficients $b_{i,j}$ contain all the terms that are finite as $n\rightarrow 0$, it remains that the decomposition between $\gamma$ and $b_{i,j}$ is not unique.

\begin{itemize}
\item We can solve eq. \eqref{eq:anom_dim_eq_resum} as a perturbative expansion in $n$, where we write the N$^k$LL anomalous dimension as a function of $\gamma_0$, the solution which determines the leading log anomalous dimension, which is $O(1)$ in the log counting:
{\small\begin{align}\label{eq:resummed_DGLAP_part}
   \gamma&=\gamma_0\Big(\frac{a_s}{n^2}\Big)+\sum_{i=1}^{\infty}n^{i}\gamma_{i}(\gamma_0)\,\\
    \gamma_0\Big(\frac{a_s}{n^2}\Big)&=\frac{1}{2}\Big(-1+\sqrt{1+4\Gamma\frac{a_s}{n^2}}\Big)\,,\\
    \gamma_{1}&=\gamma_0\Big(\frac{1+\gamma_0}{1+2\gamma_0}\Big)\Big(a_{1,0}+a_{1,1}\gamma_0\Big)\,,\\
    \gamma_{2}&=\gamma_0\Big(\frac{1+\gamma_0}{1+2\gamma_0}\Big)\Big(a_{2,0}+a_{2,1}\gamma_0+a_{2,2}\gamma_0^2\Big)\nonumber\\
    &\qquad+\frac{\gamma_0^2}{1+\gamma_0}\Big(\frac{1+\gamma_0}{1+2\gamma_0}\Big)^3\Big(a_{1,0}+\gamma_0 a_{1,1}\Big)\Big(-a_{1,0}+(1+\gamma_0) a_{1,1}\Big)\,,\\
    \gamma_{3}&=\gamma_0\Big(a_{3,0}+a_{3,1}\gamma_0+a_{3,2}\gamma_0^2+\gamma_0^3a_{3,3}\Big)\Big(\frac{1+\gamma_0}{1+2\gamma_0}\Big)\nonumber\\
    &\qquad+\frac{\gamma_0^2}{1+\gamma_0}\Big(\frac{1+\gamma_0}{1+2\gamma_0}\Big)^3\Bigg(a_{1,0}\Big(-2a_{2,0}+a_{2,1}+2a_{2,2}\gamma_0(1+\gamma_0)\Big)\nonumber\\
    &\qquad\qquad\qquad\qquad\qquad\qquad\qquad+a_{1,1}\Big(a_{2,0}+2a_{2,1}\gamma_0(1+\gamma_0)+a_{2,2}\gamma_0^2(3+4\gamma_0)\Big)\Bigg)\nonumber\\
    &\qquad-\frac{\gamma_0^3}{(1+\gamma_0)^2}\Big(\frac{1+\gamma_0}{1+2\gamma_0}\Big)^5\Big(2a_{1,0}-a_{1,1}\Big)\Big(-a_{1,0}^2+a_{1,0}a_{1,1}+\gamma_0(1+\gamma_0)a_{1,1}^2\Big)\,.
  \end{align}}
\item Matching the anomalous dimension derived from the iterations to the ansatz for the anomalous dimension equation \eqref{eq:anom_dim_eq_resum}, we can determine the coefficients $a_{k,\ell}$ in terms of $\beta_i$ and $p_{i,j,k}$:
{\small  \begin{align}\label{eq:resum_eq_determined_A}
    p_{0,-1,0}&=\frac{\Gamma}{2},\\
    a_{0,0} &= 1,\\
    a_{1,0} &= 0,\qquad a_{1,1} = -p_{0,0,0}-\frac{2}{\Gamma}\beta_0,\label{eq:trouble_maker_aij}\\
    a_{2,0} &= 0,\qquad a_{2,1} = 0,\qquad a_{2,2} = p_{0,1,0},\\
    a_{3,0} &=0,\qquad a_{3,1} = 0,\\
    a_{3,2} &=-\Big(\frac{2}{\Gamma^2}\beta_1+\frac{1}{\Gamma}\Big(-p_{1,0,0}+\beta_0 p_{0,0,1}\Big)+\frac{1}{\Gamma^2}p_{2,-2,0}\Big),\label{eq:resum_eq_determined_NcubeA}\\
    a_{3,3} &=-\Big(\frac{2}{\Gamma^2}\beta_1-p_{0,2,0}+\frac{1}{\Gamma}\Big(-p_{1,0,0}+\beta_0 p_{0,0,1}\Big)+\frac{1}{\Gamma^2}p_{2,-2,0}+\frac{p_{3,-4}}{\Gamma^3}\Big)\,.\label{eq:resum_eq_determined_NcubeB}
  \end{align}}
For compact presentation, we have dropped the contributions that are zero, like $p_{1,-2,1}$. We find for the $b_{i,j}$ coefficients:
\begin{align}
  b_{0,0} &= \frac{\Gamma}{2}p_{0,0,0}\,,\\
  b_{0,1} &= \frac{\Gamma}{2}p_{0,1,0}\,,\\
  b_{0,2} &= \frac{\Gamma}{2}p_{0,2,0}\,,\\
  b_{1,0} &= \frac{\Gamma}{2}\Bigg(p_{0,1,0}\Big(2\beta_0+\Gamma p_{0,0,0}\Big)+\Gamma p_{0,2,0}+p_{1,0,0}-\beta_0p_{0,0,1}\Bigg)\,.\label{eq:resum_eq_determined_B}
\end{align}
At this point the expansion of $\rho$ is undetermined, as given in eq. \eqref{eq:determine_rho}, and we give this in Sec. \ref{sec:finding_anomalous_dimension}.
\item The eq. \eqref{eq:generalized_resummation_equation_for_DGLAP_moment_space} implies the following resummation for the resummed coefficient function $\mathcal{R}^{T}$ defined in eq. \eqref{eq:define_coef_func_Angular_Order} to NNLL accuracy (setting $\Gamma=2$):
{\small  \begin{align}\label{eq:coefficient_function}
    \mathcal{R}^T&=\frac{1}{\sqrt{1+2\gamma_0}}+\frac{n\gamma_0^2}{(1+2\gamma_0)^{7/2}}\Bigg(p_{0,0,0}\Big(\frac{3}{2}+5\gamma_0+4\gamma_0^2\Big)+
    \beta_0\Big(\frac{1}{2}+\frac{8}{3}\gamma_0+\frac{5}{2}\gamma_0^2\Big)\Bigg)\nonumber\\
    &\quad-\frac{n^2\gamma_0(2+\gamma_0)}{4\sqrt{1+2\gamma_0}}p_{0,0,1}+\frac{n^2\gamma_0}{2(1+2\gamma_0)^{5/2}}\Big(2+11\gamma_0+16\gamma_0^2+6\gamma_0^3\Big)p_{0,1,0}-\frac{n^2\gamma_0^2}{8\sqrt{1+2\gamma_0}}p_{1,-2,1}\nonumber\\
    &\quad+\frac{n^2\gamma_0^3}{8(1+2\gamma_0)^{9/2}}\Big(-16-37\gamma_0-16\gamma_0^2+8\gamma_0^3\Big)p_{0,0,0}^2\nonumber\\
    &\quad+\frac{n^2\gamma_0^3}{12(1+2\gamma_0)^{11/2}}\Big(-4-52\gamma_0-70\gamma_0^2+31\gamma_0^3+60\gamma_0^4\Big)\beta_0 p_{0,0,0}\nonumber\\
    &\quad+\frac{n^2\gamma_0^3}{72(1+2\gamma_0)^{13/2}}\Big(-24-63\gamma_0-120\gamma_0^2+46\gamma_0^3+408\gamma_0^4+297\gamma_0^5\Big)\beta_0^2\nonumber\\
  \end{align}}
To make the result more compact, we have dropped terms originating from the $p_{i,j,k}$ that are zero, except $p_{1,-2,1}$. Reasons for this are outlined in Sec. \ref{sec:checking_coef}.
\end{itemize}
Once the coefficients to the ansatz are fixed, one can expand the ansatz to arbitrarily high order and compare against eq. \eqref{eq:iterative_expansion_fin} in the limit $R\rightarrow 0$. In practice we take the expansion to 8 or 9 loops, such that the ansatz is well over-determined.
\subsection{Comparison to Literature}
\subsubsection{The Anomalous Dimension}\label{sec:finding_anomalous_dimension}
In ref. \cite{Kom:2012hd}, the resummation of the anomalous dimension was carried out to N$^3$LL accuracy in the $\overline{MS}$ scheme by different methods.  We find to reproduce their results, we must take: 
\begin{align}\label{eq:resummation_numbers_start}
  \Gamma&=2\,,\\
  a_{0,0}&=1\,,\\
  a_{1,1}&=\frac{11}{12}\,,\label{eq:resummation_numbers_trouble}\\
  a_{2,2}&=\frac{67}{36}-\frac{\pi^2}{6}\,,\\
  a_{1,0}&=a_{2,0}=a_{2,1}=0\,,\\\label{eq:resummation_numbers_end}
  a_{3,0}&=a_{3,1}=0\,,\\
  a_{3,2}&=\frac{9}{8}\zeta_3+\frac{55\pi^2}{288}-\frac{1019}{216}\,,\\
  a_{3,3}&=\frac{3}{8}\zeta_3+\frac{55\pi^2}{288}-\frac{101}{36}\,.
\end{align}
Note that this is in complete agreement with eqs. \eqref{eq:resum_eq_determined_A} to \eqref{eq:resum_eq_determined_NcubeB}, using the relations given in eqs. \eqref{eq:gamma_s_to_p_start} to \eqref{eq:gamma_s_to_p_end}, and substituting in the constants from App. \ref{sec:constants}. Further, we find that we must take:
\begin{align}
\rho&=\beta_0\frac{\alpha_sC_A}{\pi}+\beta_1\Big(\frac{\alpha_sC_A}{\pi}\Big)^2+...
\end{align}

Alternatively, rather than comparing to ref. \cite{Kom:2012hd}, the determination of $\rho$ and the determination of the $p_{i,j,k}$ terms that contribute to $a_{i,j}$ with $i\leq 2$ follows from the leading order and next-to-leading order calculation of the time-like anomalous dimension, found in ref. \cite{Furmanski:1980cm}.

\subsubsection{Coefficient Function for $\phi\rightarrow h+X$}\label{sec:checking_coef}
In ref. \cite{Kom:2012hd}, the coefficient function for the $\phi\rightarrow h+X$ process was determined to N$^2$LL accuracy, and we can compare to their results. $\phi$ is a color singlet scalar interacting with the gauge theory through a coupling $\phi F^2$. This requires knowledge of the order $\epsilon$ terms like $p_{0,0,1}$. These contributions are uniquely determined by the space-like DGLAP anomalous dimension in $4-2\epsilon$ dimensions, using the relation of eq. \eqref{eq:fixing_angular_ordering_kernel}, and their contribution to the coefficient function is universal for any process that includes gluon fragmentation, which allows us to calculate the process dependent contribution to the process $\phi\rightarrow h+X$. We note the factorization of the $\phi\rightarrow h+X$ form factor, eq. \eqref{eq:factor_angular_frag_func} and \eqref{eq:define_coef_func_Angular_Order} after $R\rightarrow 0$, takes the form in moment space:
\begin{align}
xD_{\phi}(x)&=\int\displaylimits_{c-i\infty}^{c+i\infty}\frac{dn}{2\pi i}x^{-n}\bar{\tilde{C}}^{T}_{\phi}\Big(n,\alpha_s\Big)\mathcal{R}^T\Big(n,\alpha_s\Big)\text{exp}\Big(\int_{0}^{\alpha_s}\frac{d\alpha}{\beta(\alpha,\epsilon)}\gamma^T(\alpha,n)\Big)\,.
\end{align}  
All process dependence is now factored into the function $\bar{\tilde{C}}^{T}_{\phi}$, at least to $\NNLL$ accuracy. The function $\mathcal{R}^T$ is given in eq. \eqref{eq:coefficient_function}, and the $p_{i,j,k}$ are fixed via eq. \eqref{eq:fixing_angular_ordering_kernel} and eqs. \eqref{eq:gamma_s_to_p_start} to \eqref{eq:gamma_s_to_p_end}. Then using the fixed order result for the process $\phi\rightarrow h+X$ given in ref. \cite{Almasy:2011eq}, and the constants of App. \ref{sec:constants}, and we have the matching:
\begin{align}
  p_{0,0,1}&=-\frac{389}{72}-\frac{\pi^2}{6}+y\,,\\
  p_{1,-2,1}&=-\frac{389}{36}+2y\,,\\
\bar{\tilde{C}}^{T}_{\phi}\Big(n,\alpha_s\Big)&=1+\frac{\alpha_s C_A}{\pi}y\,.
\end{align}  
$y$ here is an arbitrary real number. We note also that for any value of $y$, this assignment agrees with the resummed results of ref. \cite{Kom:2012hd}. Using the BFKL eq. at higher orders in $\epsilon$, we can fix $p_{1,-2,1}=0$ via eq. \eqref{eq:fixing_angular_ordering_kernel}, and thus we conclude:
\begin{align}
  p_{0,0,1}&=-\frac{\pi^2}{6}\,,\\
  p_{1,-2,1}&=0\,,\\
\bar{\tilde{C}}^{T}_{\phi}\Big(n,\alpha_s\Big)&=1+\frac{\alpha_s C_A}{\pi}\frac{389}{72}\,.
\end{align}  

\section{Comparing Celestial BFKL to Angular-Ordered DGLAP}\label{celestial_BFKL_versus_angular_DGLAP}
At three loop order, for the $\NNNLL$ contribution to the resummation of the coefficient function, we have a potential conflict in the relation of eq. \eqref{eq:celestial_BFKL_to_Frag}. If we strictly define the fragmentation function $\bar{d}$ in accordance to eq. \eqref{eq:generalized_resummation_equation_for_DGLAP_momentum_space}, then the coefficient function will contain the resummation that is tied to only the structure of the $4-2\epsilon$ dimensional space-like DGLAP kernel, omitting any contribution from the fluctuation factor that plays a role in the resummation of the coefficient function in the BFKL theory for DIS. To see this, we write out the contribution to the coefficient function from the celestial BFKL eq. \eqref{eq:BFKL_evolution_small_angle} at $\NNNLL$ order:
{\small\begin{align}
R^T\Big|_{\NNNLL}&=\frac{8}{3n^3}\zeta_3\Big(\frac{\alpha_s C_A}{\pi}\Big)^3-\frac{94}{3 n^5}\zeta_3\Big(\frac{\alpha_s C_A}{\pi}\Big)^4+\frac{908}{3 n^7}\zeta_3\Big(\frac{\alpha_s C_A}{\pi}\Big)^5-\frac{8188}{3 n^9}\zeta_3\Big(\frac{\alpha_s C_A}{\pi}\Big)^6+...  
\end{align}}
Note that this will not be the full determination of the resummation of the coefficient calculation at $\NNNLL$ order, but only the contribution from the leading order celestial BFKL equation, which does not have a well-defined logarithmic order in the time-like case. At three loop order, we can write down the full contribution that arises from the angular-ordered DGLAP equation (eq. \eqref{eq:define_coef_func_Angular_Order}), which incorporates all contributions from the $4-2\epsilon$ dimensional anomalous dimension:
{\small\begin{align}
  \mathcal{R}^T\Big|_{\alpha_s^3,\NNNLL}&=\Big(\frac{\alpha_s C_A}{\pi n}\Big)^3\Bigg(\frac{4}{3}\beta_1-\frac{7}{3}\beta_0 p_{0,-1,2}-\frac{14}{3}\beta_0 p_{0,0,1}-\frac{26}{3}p_{0,0,0}p_{0,0,1}+\frac{28}{3}\beta_0p_{0,1,0}\nonumber\\
  &\qquad+4p_{0,0,0}p_{0,1,0}-\frac{34}{3}p_{0,1,1}+16p_{0,2,0}+\frac{7}{3}p_{1,-1,1}+\frac{16}{3}p_{1,0,0}-\frac{1}{3}p_{2,-3,1}-\frac{4}{3}p_{2,-2,0}\Bigg)\,.
\end{align} }
We have dropped terms which are zero as recorded in App. \ref{sec:constants}. The only contribution to the coefficient function originating from the anomalous dimension in the space-like case at this loop order, has the form:
\begin{align}
\mathcal{R}^S=1-\Big(\frac{\alpha_s C_A}{\pi n}\Big)^3\frac{p_{2,-3,1}}{3}+...\,.
\end{align}  
This allows us to extract the fluctuation factor contribution:
\begin{align}
\mathcal{N}^S&=1+2\zeta_3\Big(\frac{\alpha_s C_A}{\pi n}\Big)^3+...\,.
\end{align}  
So long as the coefficient function in the time-like case does not receive any contribution from the NLO and NNLO celestial BFKL equation, other than that captured by the angular-ordered DGLAP evolution equation and the $4-2\epsilon$ anomalous dimension, we can hazard the guess that the full contribution to the $\NNNLL$ logs at three-loop order will be given by:
\begin{align}
R^T\Big|_{\alpha_s^3,\NNNLL}&=\mathcal{R}^T\Big|_{\alpha_s^3,\NNNLL}+2\zeta_3\Big(\frac{\alpha_s C_A}{\pi n}\Big)^3\,.
\end{align}  
This is just the intuition that the time-like fluctuation factor and the space-like fluctuation factor coincide at three-loop order for the $n^{-3}$ poles. This need not be true, and hinges on the exact contributions from the NLO and NNLO celestial BFKL equations, and whether these contributions solely enter through the $4-2\epsilon$ anomalous dimension.

The only robust prediction that we can make assuming the continued validity of the celestial BFKL equation is that the $\NNNLL$ contributions to the coefficient function will not be uniquely captured by the $4-2\epsilon$ anomalous dimension, and so the relation of eq. \eqref{eq:celestial_BFKL_to_Frag} must be modified. A calculation at three-loop order for the $\phi\rightarrow h+X$ SIA process could potentially clear the situation, however, one would also have to calculate the matching coefficient \eqref{eq:factor_angular_frag_func} at this logarithmic order, and rule out any contribution from it that obscures the contribution from the celestial BFKL equation. Ultimately, one may be faced needing to perform calculations from the NLO and NNLO celestial BFKL equations and the 4-loop coefficient function to verify or rule out the fluctuation factor contribution, and thus the whole celestial BFKL approach.

\section{Towards Full Flavor QCD}
Certain generalizations are easy to guess. Adding in flavor structure to the anomalous dimension for the angular-ordered evolution is straightforward: the fragmentation functions become vectors and the kernels for angular evolution become matrices. The factorization for the form factor would be written as:
\begin{align}\label{eq:factor_angular_frag_func_flavor}
xD(x,R^2,Q^2)&=\sum_{a}\int_{x}^{1}\frac{dz}{z}\frac{x}{z}d_a\Big(\frac{x}{z},R^2,R_f^2,\frac{\mu^2}{Q^2}\Big)z\tilde{C}^{T}_a\Big(z,R_f^2,\frac{\mu^2}{Q^2}\Big)\,.
\end{align}  
The index $a$ denotes the flavor of the parton that initiates the fragmentation process. Setting $R_f=1$ and suppressing the dependence of the various functions on it, the angular-ordered evolution equation now takes the form:
{\small\begin{align}\label{eq:generalized_resummation_equation_for_DGLAP_momentum_space_flavor}
    R^2\frac{\partial}{\partial R^2} x^{1+2\epsilon} d_{a}\Big(x,R,\frac{\mu^2}{Q^2}\Big)&=\sum_{b}\rho_{ab}\Big(\frac{\mu^2}{R^2Q^2}\Big)x^{1+2\epsilon} d_{b}\Big(x,R,\frac{\mu^2}{Q^2}\Big)\nonumber\\
    &\qquad+\sum_{b}\int_{x}^{1}\frac{dz}{z}P_{ab}\Big(\frac{x}{z};\frac{\mu^2}{z^2R^2Q^2}\Big)z^{1+2\epsilon} d_{b}\Big(z,R,\frac{\mu^2}{Q^2}\Big)\,,\\
  P_{ab}\Big(\frac{x}{z};\frac{\mu^2}{z^2R^2Q^2}\Big)&=\sum_{\ell=1}^{\infty} P^{(\ell-1)}_{ab}\Big(\frac{x}{z};a_s;\epsilon\Big)\,\Big(\frac{\mu^2}{z^2R^2Q^2}\Big)^{\ell\epsilon}\,,\\
  \rho_{ab}\Big(\frac{\mu^2}{R^2Q^2}\Big)&=\sum_{\ell=1}^{\infty}\rho^{(\ell-1)}_{ab}(a_s;\epsilon)\Big(\frac{\mu^2}{R^2Q^2}\Big)^{\ell\epsilon}\,.
\end{align}}
While the reciprocity relation of eq. \eqref{eq:reciprocity} is not straightforward to generalize as a matrix equation, but we can rewrite eq. \eqref{eq:fixing_angular_ordering_kernel} easily enough:
\begin{align}
  \int_0^{1}\frac{dR^2}{R^2}\Big(\rho_{ab}(R^{-2})+\bar{P}_{ab}(n,\epsilon,R^{-2})\Big)&=\int_{0}^{\alpha_s}\frac{d\alpha}{\beta(\alpha,\epsilon)}\gamma^u_{ba}(\alpha,n,\epsilon)\,,
\end{align}
where $a,b$ are flavor indicies. In the same spirit as the reciprocity relation, this implies the underlying time-like anomalous dimension is still wholely determined by the $4-2\epsilon$ dimensional space-like DGLAP kernel with arbitrary flavors which can be constructed by consistency with the BFKL equation. 

Whether or not the resummed time-like anomalous dimension in the full flavor case could be captured by a matrix equation generalization of eq. \eqref{eq:anom_dim_eq_resum} remains to be seen, as well as whether $\rho_{ab}$ can be straightforwardly identified with the beta function.

\section{Discussion}

We have introduced two different ways to tackle the resummation of soft effects in dimensional regularization for fragmentation. The first way is through the celestial BFKL equation of Sec. \ref{sec:BMS_BFKL}, while the second is the angular-ordered evolution equation in dimensional regularization of Sec. \ref{sec:conjectures}. The second approach reproduces results to $\NNNLL$ in the anomalous dimension and $\NNLL$ in the coefficient function, and the two approaches are consistent where we can easily check, LL in the coefficient function, and partial results to all subleading logarithmic orders in the anomalous dimension, limited only by our ability to calculate.

The two approaches are complementary: the celestial BFKL orders in energy, and emissions can occur at any angle, while the angular-ordered evolution in dimensional regularization orders in angle, and emissions can occur at any energy. We have attempted to connect the two with eq. \eqref{eq:fixing_angular_ordering_kernel}, where we directly tie the angular-ordered kernel to the anomalous dimension determined by BFKL theory in $4-2\epsilon$ dimensions. Logically, it seems possible that the celestial BFKL equation fails at some order, perhaps due to lack of underlying conformal symmetry in QCD, while the angular-ordered equation remains valid to all orders. Failure of the angular-ordered evolution equation does not rule out the celestial BFKL approach: indeed, it may explain the origin of terms at $\NNNLL$ in the coefficient function that the angular-ordered evolution misses. We attempt to outline calculations that could help clarify the matter in Sec. \ref{celestial_BFKL_versus_angular_DGLAP}.

We have as of yet no direct way to calculate the angular-ordered fragmentation function, since we do not have a matrix element definition of the fragmentation function. We can recourse to the postulated relationship of eq. \eqref{eq:fixing_angular_ordering_kernel}, but since the BFKL equation becomes increasingly unwieldy at higher orders, it would be desirable to have a direct means to calculate higher order terms in $\epsilon$ of the anomalous dimension at fixed order, independent of the underlying process. Otherwise, comparing directly to a fixed order calculation of a specific process we see there can arise ambiguities in the determination of the $\epsilon$ expansion of the anomalous dimension, like that discussed in Sec. \ref{sec:checking_coef}. 

Finally, we also note that we have evolved the small angle limit of eq. \eqref{eq:BFKL_for_Frag_One_Loop}. For the anomalous dimension, which is sensitive to the collinear singularity structure, this should be sufficient. Indeed, it seems likely this ought to be sufficient for the problem of fragmentation, which is driven by the collinear dynamics. However, one should check explicitly that solving the large angle celestial BFKL equation produces no additional contributions versus its small angle expansion, that these contributions are genuinely higher ``twist'' with respect to leading order factorization with the fragmentation function.

\section{Conclusions}

A chief conceptual result of the paper is a new way to understand the effectiveness of angular ordering in QCD evolution equations. Ultimately, we claim that the angular ordering of the time-like parton shower is the result of the map between BK/BFKL/JIMWLK theory and the theory of eikonal lines extending out to the celestial sphere described in the BMS equation. Under this map, angles on the celestial sphere found in jet physics map to positions in the transverse plane to the colliding beams in DIS. Large transverse momentum transfers that probe the PDF in DIS localize the dynamics in a small spatial region of the transverse plane. This spatial region is set by the transfered momentum in the t-channel exchange, and the DGLAP evolution equation in DIS describes how the cross-section changes as one varies the size of the probed region. This effective cutoff in position space maps to a cutoff in angles between eikonal lines on the celestial sphere, up to anomalous dependence on an energy scale. Varying the momentum scale in DIS maps to varying the angles in SIA. The wrinkle is the mismatch in units between positions and angles. Strictly in 4 dimensions, in a conformal theory, the mismatch between the dimensionless angles and the dimensionful positions does not matter. But to calculate the anomalous dimension, the theory must be regulated, so that even in a conformal theory, one needs the appearance of the energy scale in the fragmentation process to make up the mismatch. This leads to the different structure of the time-like versus space-like anomalous dimensions, and ultimately, the difference in the fragmentation spectrum versus the PDF. Ultimately, the most pleasing result is a deeper understanding of the old problem of Drell-Levy-Yan, the correct way to analytically continue between DIS and SIA: one should map angles to positions, paying careful attention to the regularization scheme.  

The advantage to the resummation scheme of the time-like anomalous dimension and coefficient functions presented in ref. \cite{Kom:2012hd} is a more thorough notion of factorization in the soft and collinear limits. We have a precise recipe for determining the universal contribution to the matching coefficients, and their resummation. Thus we can factorize out process dependent contributions, hopefully opening the door to a resummation of soft fragmentation with \emph{hadronic} initial states, $e+p\rightarrow h+X$ and $p+p\rightarrow h+X$, or the resummation of semi-inclusive jet production of refs. \cite{Kang:2016mcy,Dasgupta:2016bnd,Dai:2016hzf}. The ability to tackle hadronic initial states and semi-inclusive jet production would be useful for comparing to the fixed order results of refs. \cite{Currie:2016bfm,Currie:2018xkj,Czakon:2019tmo}. 

Pushing the BFKL theory for fragmentation will be a more daunting task than the angular-ordered dimensionally regulated DGLAP equation, if experience with space-like BFKL theory is any indication (see, e.g., \cite{Ball:2005mj,Ciafaloni:2005cg,Ciafaloni:2006yk}). The most difficult part may be understanding the difference in log-counting: the angular-ordered dimensionally regulated DGLAP equation has manifest log counting, while the BFKL approach mixes orders in the time-like case. What remains to be seen is whether the BFKL theory can capture corrections to the coefficient functions not achieved in the DGLAP approach, or whether one or the other or both approaches need a modification of the initial conditions to the evolution in such a way as to generate the correct time-like coefficient functions. Moreover, beyond LO, we would need to check that the collinear expansion of the celestial BFKL equation suffices to overcome the lack of conformal invariance in a general gauge theory, which spoils the BMS/BFKL correspondence. 

Another fascinating possibility would be to formulate a time-like effective theory that is the direct counterpart to the forward scattering effective theory developed in ref. \cite{Rothstein:2016bsq}. The results of this paper lend creedence to the idea that the Glauber lagrangian of ref. \cite{Rothstein:2016bsq} should have a time-like counterpart, and one should be able to map the rapidity factorization of BFKL theory to a energy-ordered factorization of jet processes on the celestial sphere governed by an oxymoronic ``time-like potential mode.'' In rapidity factorization one can motivate leaving the evolution equation in $4-2\epsilon$ dimensions, since the counter-terms for rapidity divergences can be defined to all orders in $\epsilon$ (ref. \cite{Chiu:2012ir}). Thus deriving a forward scattering BFKL equation in $4-2\epsilon$ is possible in the framework of ref. \cite{Rothstein:2016bsq}. We do not encounter rapidity divergences in the celestial BFKL equation. Dimensional regularization handles both divergences in energy and angular integrals. Since we do not take $\epsilon$ to zero until after we evolve, we have no renormalization group procedure (except perhaps in a Wilsonian-cutoff sense) to derive the celestial BFKL equation, which may require revisiting the approach of ref. \cite{Becher:2016mmh} which uses renormalization in dimensional regularization for a unified theory of scattering dynamics on the celestial sphere. 

\section{Acknowledgements}
D.N. wishes to thank Iain Stewart for many conversations about the resummation of DIS as $x\rightarrow 0$ that ultimately laid the groundwork for this paper, Wouter Waalewijn for collaboration on fragmentation, and to Ian Moult for conversations on the reciprocity relations and reading the manuscript. D.N. was supported by the Department of Energy under Contract DE-AC52-06NA25396 at LANL and through the LANL/LDRD Program via a Feynman Distinguished Fellowship. F.R. was supported by the U.S.~Department of Energy under Contract No.~DE-AC02-05CH11231, the LDRD Program of Lawrence Berkeley National Laboratory, the National Science Foundation under Grant No.~ACI-1550228 within the JETSCAPE Collaboration.

\appendix
\section{Constants}\label{sec:constants}
We write for the space-like anomalous dimension:
\begin{align}
  \gamma^{S}(n)&=\sum_{i=0}^{\infty}\sum_{j=-i-1}^{\infty}a_s^{i+1}n^{j}\gamma^{s}_{i,j}\,,\\
  a_s&=\frac{\alpha_sC_A}{\pi}\,.
\end{align}  
In momentum space, we write:
\begin{align}
  \gamma^{S}(x)&=a_s \gamma^{S(0)}(x)+a_s^2 \gamma^{S(1)}(x)+...\\
  \gamma^{T}(x)&=a_s \gamma^{T(0)}(x)+a_s^2 \gamma^{T(1)}(x)+...
\end{align}  
With our normalization rules, we have:
\begin{align}
\beta_0&=\frac{11}{12}\,,\\
\beta_1&=\frac{17}{24}\,,\\
\gamma^{s}_{0,0}&=-\frac{11}{12}\,,\\
\gamma^{s}_{0,1}&=\frac{67}{36}-\frac{\pi^2}{6}\,,\\
\gamma^{s}_{0,2}&=-\frac{413}{216}+\zeta_3\,,\\
\gamma^{s}_{1,0}&=\frac{1643}{216}-\frac{11}{36}\pi^2-2\zeta_3\,,\\
\gamma^{s}_{2,-2}&=-\frac{395}{108}+\frac{11}{72}\pi^2+\frac{\zeta_3}{2}\,,\\
\gamma^{s}_{3,-4}&=2\zeta_3\,,\\
\gamma^{s}_{1,-2}&=\gamma^{s}_{1,-1}=\gamma^{s}_{2,-3}=0\,.
\end{align}
Since the two-loop time-like anomalous dimension is explicitly known from direct calculation (ref. \cite{Furmanski:1980cm}), we can perform so checks of our resummation. Using the map of eqs. \eqref{eq:gamma_s_to_p_start} to \eqref{eq:gamma_s_to_p_end}, and eq. \eqref{eq:resum_eq_determined_B}, we can compare against the explicit calculation of the integral of the time-like anomalous dimension, subtracting the singular terms as $x\rightarrow 0$. We then get:
\begin{align}
  \int dx \Big\{\gamma^{T(1)}(x)-\Big(\gamma^{T(1)}(x)\Big|_{x\rightarrow 0}\Big)\Big\}&=\frac{10}{27}\nonumber\\
  &=2\gamma^{s}_{0,0}\gamma^{s}_{0,1}+2\gamma^{s}_{0,2}+\gamma^{s}_{1,0}
\end{align}  
We note we also reproduce from eqs. \eqref{eq:resum_eq_determined_NcubeA} and  \eqref{eq:resum_eq_determined_NcubeB}:
\begin{align}
  a_{3,2}&=\frac{9}{8}\zeta_3+\frac{55\pi^2}{288}-\frac{1019}{216}=-\frac{1}{2}\gamma^{s}_{1,0}+\frac{1}{4}\gamma^{s}_{2,-2}\\
  a_{3,3}&=\frac{3}{8}\zeta_3+\frac{55\pi^2}{288}-\frac{101}{36}=-\gamma^{s}_{0,2}-\frac{1}{2}\gamma^{s}_{1,0}+\frac{1}{4}\gamma^{s}_{2,-2}+\frac{1}{8}\gamma^{s}_{3,-4}
\end{align}  
We also need the higher order terms in the $\epsilon$-expansion of $\gamma^u$, which determine the $p_{i,j,k}$ coefficients. We can obtain the expansion of $\gamma^u$ via the $4-2\epsilon$ dimensional BFKL equation, using the results of ref. \cite{Ciafaloni:2005cg}. This determines all coefficients of the form $p_{i,-i-1,k}$, for $i,k\geq 0$. The needed coefficients are:
\begin{align}
  p_{0,0,1}&=-\frac{\pi^2}{6},\\
  p_{1,-2,1}&=p_{0,-1,1}=0\,,\\
  p_{2,-3,1}&=-2\zeta_3\,\\
  p_{0,-1,2}&=-\frac{\pi^2}{12},\\
  p_{1,-2,2}&=-2\zeta_3\,.
\end{align}

\section{Stereographic Mapping and BFKL}\label{sec:stereographics}
First we review the mapping in strictly 2 dimensions. The mapping is the stereographic projection of the celestial sphere to the transverse plane. We let ``north'' on the celestial sphere point along the spatial direction of the light-cone direction $n$, and the south along $\bar{n}$. The sphere is tangent to the plane at the point $\hat{n}$. We note under the conformal mapping relating space-like to time-like evolution:
\begin{align}\label{eq:map_angles_to_distances}
n_a\cdot n_b&=1-\cos\theta_{ab}=\frac{2(\vec{x}_{\perp a}-\vec{x}_{\perp b})^2}{(1+\vec{x}_{\perp a}^2)(1+\vec{x}_{\perp b}^2)}\,.
\end{align}
Coordinates are mapped as:
\begin{align}
\cos\theta =\frac{1-\vec{x}^2_{\perp}}{1+\vec{x}^2_{\perp}}\,,\,\sin\theta =\frac{2|\vec{x}_\perp|}{1+\vec{x}^2_\perp}\,,&\,\sin\theta\cos\phi =\frac{2x_1}{1+\vec{x}^2_\perp}\,,\,\sin\theta\sin\phi =\frac{2x_2}{1+\vec{x}^2_\perp}\,,\\
\vec{x}^2_\perp&=\tan^2\frac{\theta}{2}=\frac{n\cdot n_\theta}{\bar{n}\cdot n_\theta}\,,\\
\vec{x}&=(x_1,x_s)
\end{align}
The vector $\vec{x}_\perp$ is a point in the transverse plane we map the sphere to. Finally:
\begin{align}
\frac{d^2\Omega_j}{(1+\cos\theta_j)^2}&=d^2\Omega_j\Big(\frac{1+\tan^2\frac{\theta_j}{2}}{2}\Big)^2=d^2\vec{x}_{\perp j},\\
\int d^2\Omega_j\frac{n_a\cdot n_b}{(n_a\cdot n_j)(n_j\cdot n_b)}&=2\int d^2\vec{x}_{\perp j}\frac{(\vec{x}_{\perp a}-\vec{x}_{\perp b})^2}{(\vec{x}_{\perp a}-\vec{x}_{\perp j})^2(\vec{x}_{\perp j}-\vec{x}_{\perp b})^2}\,.
\end{align}
Now let $X_i$ by the cartesian coordinates of a $d$-dimensional space where we embed a $d-1$-sphere. Let $\vec{x}_\perp=(x_1,...,x_{d-1})$ be the coordinates of the hyperplane we are mapping to. We have for the stereographic projection:
\begin{align}
  X_i&=\frac{2x_i}{1+\vec{x}_\perp^{\,2}},\, \text{ if } i=1,...,d-1\,,\\
  X_d&=\frac{1-\vec{x}_\perp^{\,2}}{1+\vec{x}_\perp^{\,2}}=\cos\theta\,.
\end{align}  
These $X_i$ coordinates define a valid point on the sphere. Since relation eq. \eqref{eq:map_angles_to_distances} still holds in $d=2-2\epsilon$ \emph{spatial} dimensions, then eikonal integrals under the stereographic projection map as:
\begin{align}
\int d^{2-2\epsilon}\Omega_j\Big(\frac{1+\text{tan}^2\frac{\theta_j}{2}}{2}\Big)^{-2\epsilon}\frac{n_a\cdot n_b}{(n_a\cdot n_j)(n_j\cdot n_b)}=2\int d^{2-2\epsilon}\vec{x}_{\perp j}\frac{(\vec{x}_{\perp a}-\vec{x}_{\perp b})^2}{(\vec{x}_{\perp a}-\vec{x}_{\perp j})^2(\vec{x}_{\perp j}-\vec{x}_{\perp b})^2}\,.
\end{align}  
The spatial direction of the light-cone vector $n$ (the jet direction) defines the $d$-th coordinate. We are interested in the region where the angular cutoff $R\rightarrow 0$, and thus we see that expanding in the small angle limit $\theta_j,\theta_a,\theta_b\sim R$:
\begin{align}
\int d^{2-2\epsilon}\Omega_j\frac{n_a\cdot n_b}{(n_a\cdot n_j)(n_j\cdot n_b)}+...=2^{1+2\epsilon}\int d^{2-2\epsilon}\vec{x}_{\perp j}\frac{(\vec{x}_{\perp a}-\vec{x}_{\perp b})^2}{(\vec{x}_{\perp a}-\vec{x}_{\perp j})^2(\vec{x}_{\perp j}-\vec{x}_{\perp b})^2}\,.
\end{align}  

\section{Reciprocity Equations}\label{sec:reciprocity}
We review how one can obtain the small-x resummation of the time-like anomalous dimension from the reciprocity relations of refs. \cite{Dokshitzer:2005bf,Basso:2006nk}. We introduce the reciprocity kernel $P_r$, which governs both the space-like and time-like anomalous dimensions as follows:
\begin{align}\label{eq:reciprocity_alt}
  \bar{P}_r&\Big(n+2\gamma^T(n)+2\beta_0a_s+2\beta_1 a_s^2+...\Big)=\gamma^T(n)\,,\\
  \bar{P}_r&\Big(n-2\gamma^S(n)-2\beta_0a_s-2\beta_1 a_s^2+...\Big)=\gamma^S(n)\,,\\
  \bar{P}_r(n)&=\sum_{m=0}^{\infty}\bar{P}^{(m)}_r(n)\,.\label{eq:universal_anom_dimension}
\end{align}  
$\bar{P}_r$ is the reciprocity kernel. We expand as:
\begin{align}%
\gamma^{T}(n)&=\frac{n}{2}\Big(\gamma_0\Big(\frac{a_s}{n^2}\Big)+n h_1(\gamma_0)+n^2 h_2(\gamma_0)+n^3 h_3(\gamma_0)+...\Big)+\sum_{i=0}^{\infty}\sum_{j=0}^{\infty} a_s^{i+1} n^j b_{i,j}\\      
\bar{P}_r(n)&=a_s\Bigg(\frac{1}{n}+p_{0,0}^{r}+np_{0,1}^{r}+n^2p_{0,2}^{r}+...\Bigg)+a_s^2p_{1,0}^{r}+\frac{a_s^3}{n^2}p_{2,-2}^{r}+\frac{a_s^4}{n^4}p_{3,-4}^{r}+...\\
a_s&=\frac{\alpha_s C_A}{\pi}
\end{align}
We use the same ansatz for the time-like anomalous dimension as eq. \eqref{eq:anom_dim_eq_resum}, and the same normalization as eq. \eqref{eq:gamma_def}. And we expand the anomalous dimensions as before, obtaining the $a_{i,j}$ and $b_{i,j}$ coefficients defined in those eqs.:
{\small\begin{align}
a_s\Gamma&=n^2\gamma_0(1+\gamma_0),\\
\Gamma &=2,\\
a_{1,0}&=0,\qquad a_{1,1}=-(p_{0,0}^{r}+\beta_0),\\
a_{2,0}&=0,\qquad a_{2,1}=0,\qquad a_{2,2}=p_{0,1}^{r},\\
a_{3,0}&=0,\qquad a_{3,1}=0,\\
a_{3,2}&=-\Big(\frac{\beta_1}{2}+\frac{1}{2}p_{1,0}^{r}-\frac{1}{4}p_{2,-2}^{r}\Big),\\
a_{3,3}&=-\Big(\frac{\beta_1}{2}+p_{0,2}^{r}+\frac{1}{2}p_{1,0}^{r}-\frac{1}{4}p_{2,-2}^{r}-\frac{1}{8}p_{3,-4}^{r}\Big),\\
b_{0,0}&=p_{0,0}^{r},\\
b_{0,1}&=p_{0,1}^{r},\\
b_{0,2}&=p_{0,2}^{r},\\
b_{1,0}&=2\Bigg(p_{0,2}^{r}+(p_{0,0}^{r}+\beta_0)p_{0,1}^{r}+\frac{1}{2}p_{1,0}^{r}\Bigg).
\end{align}}
We then find that we can map reciprocity kernel's coefficients to the space-like anomalous dimension:
\begin{align}
p_{i,j}^r&=\gamma^S_{i,j},\, \text{ if } j\neq 0\,,\\
p_{i,0}^r&=\gamma^S_{i,0}+\beta_i.\,
\end{align}

\section{Derivation of Celestial BFKL}\label{sec:derivation}
We construct the probability to find $k+2$ emissions with energy greater than $xQ$ emitted from a dipole with directions $n_a$ and $n_b$. We include the dipole legs when counting emissions. To do so, we use the leading logarithmic form of soft gluon cross-section (see ref. \cite{Banfi:2002hw} and its references):
{\small\begin{align}
    P_{k+2}(x,n_a\cdot n_b)&=(4\pi\alpha_s C_A)^n\mu^{2\epsilon k}e^{k\epsilon\gamma_E}(4\pi)^{k\epsilon}\int\prod_{i=1}^k\frac{d^{2-2\epsilon}\Omega_i}{(2\pi)^{3-2\epsilon}}\frac{d\omega_i}{2\omega^{1+2\epsilon}_i}\Theta\Big(\omega_i-Qx\Big)W_{ab}(1,2,...,k)  ,\\
    W_{ab}(1,2,...,k)&=\frac{n_{a}\cdot n_b}{(n_{a}\cdot n_1)\,(n_{1}\cdot n_2)\,...\,(n_{k}\cdot n_b)}\,.
\end{align}}
The virtual corrections which we have omitted will be built into the evolution equation by an appropriate subtraction. All null vectors are of the form $n_i=(1,\hat{n}_i)$, with $\hat{n}_i$ being the direction the momentum of the emission is flowing on the celestial sphere. Further, $\omega_i$ is the energy of the emission,\footnote{In a more lorentz invariant definition, if we take $Q$ to be time-like momentum of the hard-process that initiates the cascade, $\omega_i=\frac{Q\cdot p_i}{\sqrt{Q^2}}$, where $p_i$ is the momentum of the emission.} and we have solved the on-shell condition as:
\begin{align}
\int \frac{d^dp}{(2\pi)^{d-1}}\theta(p^0)\delta(p^2)&=\frac{1}{2}\int\frac{d^{2-2\epsilon}\Omega}{(2\pi)^{3-2\epsilon}}\frac{d\omega}{\omega^{1+2\epsilon}}\,.
\end{align}  
We now take the ln$x$-derivative to get:
{\small\begin{align}\label{eq:evolve_exclusive_jets}
    x\frac{d}{dx}P_{k+2}(x,n_a\cdot n_b)=\SignOnPosBFKL\frac{\alpha_sC_A}{\pi}\Big(\frac{\mu e^{\frac{\gamma_E}{2}}}{xQ}\Big)^{2\epsilon}\int\frac{d^{2-2\epsilon}\Omega_j}{4\pi^{1-\epsilon}}\frac{n_a\cdot n_b}{n_a\cdot n_j\,n_j\cdot n_b}\Bigg\{\sum_{\ell=1}^{k-1}&P_{\ell+1}\Big(x,n_a\cdot n_j\Big)P_{k-\ell+1}\Big(x,n_j \cdot n_b\Big)\nonumber\\
    &-P_{k+2}\Big(x,n_a\cdot n_b\Big)\Bigg\}\,.
\end{align}}
The subtracted term handles the virtual corrections, and we have used the trick:
\begin{align}
W_{ab}(1,2,...,k)=\frac{n_a\cdot n_b}{n_a\cdot n_\ell\,n_\ell\cdot n_b}W_{a\ell}(1,2,...,\ell-1)W_{\ell b}(\ell+1,...,k)\,.
\end{align}  
Further, we relabel $n_\ell\rightarrow n_j$. We then construct the average number of emissions with energy above $xQ$, the relevant quantity for fragmentation, by writing:
\begin{align}\label{eq:define_celestial_impact_factor}
{\mathcal N}\Big(x,n_a\cdot n_b\Big)&=\sum_{k=0}^{\infty}(k+2) P_{k+2}(x,n_a\cdot n_b)\,.
\end{align}  

We then take the $x$-derivative on both sides of eq. \eqref{eq:define_celestial_impact_factor}, using eq. \eqref{eq:evolve_exclusive_jets}, and the unitarity condition:
\begin{align}
 1=\sum_{k=0}^{\infty} P_{k+2}(x,n_a\cdot n_b)\,.
\end{align}  
This achieves:
{\small\begin{align}\label{eq:direct_celestial_bfkl}
    x\frac{d}{dx}{\mathcal N}(x,n_a\cdot n_b)&=\SignOnPosBFKL\frac{\alpha_sC_A}{\pi}\Big(\frac{\mu e^{\frac{\gamma_E}{2}}}{xQ}\Big)^{2\epsilon}\int\frac{d^{2-2\epsilon}\Omega_j}{4\pi^{1-\epsilon}}\frac{n_a\cdot n_b}{n_a\cdot n_j\,n_j\cdot n_b}\Bigg\{{\mathcal N}\Big(x,n_a\cdot n_j\Big)+{\mathcal N}\Big(x,n_j \cdot n_b\Big)-{\mathcal N}\Big(x,n_a\cdot n_b\Big)\Bigg\}\,.
  \end{align}}

This derivation establishes the essential modification that the celestial BFKL must have, anomalous dependence on the energy scale $xQ$ in $4-2\epsilon$ dimensions which is necessary for satisfying the reciprocity relation $\gamma^{S}(n+2\gamma^T(n))=\gamma^T(n)$. \emph{If} we were to set $\epsilon = 0$ \emph{and} identify:
\begin{align}
{\mathcal D}(x,n_a\cdot n_b)\sim \frac{d}{dx}{\mathcal N}(x,n_a\cdot n_b)\,,
\end{align}  
\emph{then} we would achieve eq. \eqref{eq:BFKL_for_Frag_One_Loop} in the limit $\epsilon=0$. Identifying ${\mathcal D}$ with the derivative of ${\mathcal N}$ is following the intuition that ${\mathcal D}$ describes the differential fragmentation spectrum, while ${\mathcal N}$ tracks the cumulative. Ultimately, we study eq. \eqref{eq:BFKL_for_Frag_One_Loop} with the factor $(1+2\epsilon){\mathcal D}$, since eq. \eqref{eq:BFKL_for_Frag_One_Loop} has the following property with respect to the solution of the traditional BFKL equation  (eq. \eqref{eq:BFKL}), which we have found from explicit calculations:
\begin{align}
  \text{ If } R^S=\sum_{i=0}^{\infty}c_{i}\Big(\frac{\alpha_sC_A}{\pi n}\Big)^{i}\text{ then } R^T=\sum_{i=0}^{\infty}c_{i}\Big(\frac{\alpha_sC_A}{\pi n}\Big)^{i}\Big(1+f_{i}\Big(\frac{\alpha_sC_A}{\pi n^2}\Big)\Big)\,,\nonumber\\
  \text{ where } f_{i}\Big(\frac{\alpha_sC_A}{\pi n^2}\Big)=\sum_{k=1}^{\infty}d_{ik}\Big(\frac{\alpha_sC_A}{\pi n^2}\Big)^k\,.
\end{align}  
In particular, while $c_i$ are numbers of ascending transcendentality, the coefficients $d_{ik}$ are all rational numbers. Thus we have the result:
\begin{align}
  R^{T}=R^{S}+O\Big(\frac{\alpha_s}{n^2}\Big)
\end{align}
Which is valid when $\alpha_s\sim n$, but $n\gg 1$. This is not the soft fragmentation regime.

Thus eq. \eqref{eq:BFKL_for_Frag_One_Loop} appears to give a resummed celestial ``impact factor'' which is ``most'' dual to the BFKL equation in the transverse plane. Specifically, if we take the fourier transform of the traditional BFKL equation, (eq. \eqref{eq:BFKL}), we achieve:
{\small\begin{align}
    x\frac{d}{dx}\mathcal{F}\big(x&,\vec{b}_{\perp ab}^{\,2}\big)=-\mathcal{F}\big(x,\vec{b}_{\perp ab}^{\,2}\big)\nonumber\\
    &\SignOnPosBFKL\Big(\frac{\mu e^{\frac{\gamma_E}{2}}}{Q}\Big)^{2\epsilon}\frac{\alpha_sC_A}{\pi}\int\frac{d^{2-2\epsilon}\vec{b}_{\perp j}}{2\pi^{1-\epsilon}}\frac{\vec{b}_{\perp ab}^{\,2}}{\vec{b}_{\perp aj}^{\,2}\vec{b}_{\perp jb}^{\,2}}\Big(\mathcal{F}\big(x,\vec{b}_{\perp aj}^{\,2}\big)+\mathcal{F}\big(x,\vec{b}_{\perp jb}^{\,2}\big)-\mathcal{F}\big(x,\vec{b}_{\perp ab}^{\,2}\big)\Big)\,.
\end{align}  }
Where $\vec{b}_{\perp ab}=\vec{b}_{\perp a}-\vec{b}_{\perp b}$, $\vec{b}_{\perp aj}=\vec{b}_{\perp a}-\vec{b}_{\perp j}$, and $\vec{b}_{\perp jb}=\vec{b}_{\perp j}-\vec{b}_{\perp b}$. Thus to map to eq. \eqref{eq:direct_celestial_bfkl}, we would like to adopt the following prescription in the collinear limit of angles on the celestial sphere mapping to positions:
\begin{align}\label{eq:map_BMS_to_BK}
xQ\vec{\theta}\leftrightarrow\vec{b}_\perp, \mathcal{F}\big(x&,\vec{b}_{\perp ab}^{\,2}\big)\leftrightarrow  {\mathcal D}(x,n_a\cdot n_b)\,.
\end{align}  
While this suffices to reproduce the collinear limit of eq. \eqref{eq:direct_celestial_bfkl}, the actual equation which gives solutions obeying a form of reciprocity in their solutions is eq. \eqref{eq:BFKL_for_Frag_One_Loop} with the additional factor of $-2\epsilon{\mathcal D}$ for reasons outlined above. 

\bibliographystyle{JHEP}
\bibliography{bibliography}


\end{document}